\documentclass{emulateapj}
\usepackage{natbib}
\newcommand{\Msun}{{\rm M_{\odot}}}

\newcommand{\kpc}{\, {\rm kpc}}
\newcommand{\pc}{\, {\rm pc}}
\newcommand{\kmps}{\, {\rm km \, s^{-1}}}

\slugcomment{Accepted for Publication in ApJ}

\shorttitle{Evolution of Gas Phase in the MW}
\shortauthors{Koda et al.}

\begin{document}

\title{Evolution of Molecular and Atomic Gas Phases in the Milky Way}

\author{
Jin Koda\altaffilmark{1,2},
Nick Scoville\altaffilmark{2},
Mark Heyer\altaffilmark{3}
}

\altaffiltext{1}{Department of Physics and Astronomy, Stony Brook University, Stony Brook, NY 11794-3800}
\altaffiltext{2}{California Institute of Technology, MC 249-17, 1200 East California Boulevard, Pasadena, CA 91125}
\altaffiltext{3}{Department of Astronomy, University of Massachusetts, Amherst, MA 01003}

\email{jin.koda@stonybrook.edu}

\begin{abstract}
We analyze radial and azimuthal variations of the phase balance between the molecular and atomic interstellar medium
(ISM) in the Milky Way using archival CO($J$=1-0) and HI 21cm data.
In particular, the azimuthal variations -- between spiral arm and interarm regions -- are analyzed
without any explicit definition of spiral arm locations.
We show that the molecular gas mass fraction, i.e., $f_{\rm mol}=\Sigma_{H_2}/(\Sigma_{HI}+\Sigma_{H_2})$,
varies predominantly in the radial direction:
starting from $\sim100\%$ at the center, remaining $\gtrsim50\%$ ($\gtrsim60\%$) to $R\sim 6\kpc$,
and decreasing to $\sim 10$-$20\%$  ($\sim 50\%$) at $R=8.5$ kpc when averaged over the whole disk thickness
(in the midplane).
Azimuthal, arm-interarm variations are secondary:
only $\sim 20\%$, in the globally molecule-dominated inner MW, but becoming larger, $\sim 40$-$50\%$,
in the atom-dominated outskirts.
This suggests that in the inner MW, the gas stays highly molecular ($f_{\rm mol}>50\%$)
as it goes from an interarm region, into a spiral arm, and back into the next interarm region.
Stellar feedback does not dissociate molecules much, and the coagulation and fragmentation of molecular clouds dominate
the evolution of the ISM at these radii.
The trend differs in the outskirts, where the gas phase is globally atomic ($f_{\rm mol}<50\%$).
The HI and H$_2$ phases cycle through spiral arm passage there.
These different regimes of ISM evolution are also seen
in external galaxies (e.g., the LMC, M33, and M51).
We explain the radial gradient of $f_{\rm mol}$ by a simple flow continuity model.
The effects of spiral arms on this analysis are illustrated in the Appendix.
\end{abstract}

\keywords{ISM: evolution --- ISM: atoms --- ISM: molecules --- ISM: clouds --- Galaxy: disk --- Galaxy: evolution}

\section{Introduction}
The evolution of the molecular and atomic gas phases in the Milky Way (MW) has been an issue
of debate for a long time (e.g., \citealt{Scoville:1979lg, Blitz:1980sh}; see  \citealt{Heyer:2015qy} for review).
In fact, two contradicting scenarios have been suggested for the evolution of gas across spiral arms.
The classic scenario posits a rapid phase transition from interarm HI gas to
molecular clouds (MCs)
\footnote{Following \citet{Solomon:1980xy} and \citet{Heyer:2015qy} we use the term "molecular clouds"
for all clouds (small and large) whereas "giant molecular clouds" for clouds with $>10^5\Msun$.}
in spiral arms, and back into the ionized and atomic phases
by photodissociation associated with star formation in the spiral arms \citep[e.g., ][]{Cohen:1980ve, Blitz:1980sh}.
Alternatively, the dynamically-driven scenario involves little phase transition and
suggests that the evolution is driven by the coagulation
and fragmentation of MCs
around spiral arms
-- assembling pre-existing small interarm MCs into more massive ones in spiral arms
and shearing them into smaller MCs toward the next interarm region \citep[e.g., ][]{Scoville:1979lg, Vogel:1988tp}.
The difference between the two scenarios (i.e., diffuse HI vs. dense H$_2$ in interarm regions)
is important when considering triggers and thresholds of star formation.

The classic scenario of the rapid phase transition is supported by observations of MCs
in the solar neighborhood, the Large Magellanic Cloud (LMC), and the nearby spiral galaxy M33.
Local Galactic MCs linked with OB stars appear photo-dissociated \citep[e.g., ][]{Tachihara:2001qy}.
Young star clusters often show no parent MCs nearby as the cluster ages approach
$\sim 10$-$40$ Myr, and perhaps the MCs are destroyed due to stellar feedback \citep{Hartmann:2001mk, Kawamura:2009lr, Miura:2012yq}.
Molecular gas and clouds exist primarily, within HI spiral arms with little or no H$_2$ gas in the interarm regions
\citep{Engargiola:2003jo, Heyer:2004zh, Fukui:2009lr}.
Thus, the MC lifetimes should be as short as an arm crossing time $\sim$30 Myr.
The molecular gas should therefore be dissociated after spiral arm passages.

Early interferometric observations of the grand-design spiral galaxy M51 showed
an {\it apparent} absence of molecular gas in its interarm regions
\citep{Vogel:1988tp, Rand:1990fb} since it was largely resolved out.
High-fidelity imaging of molecular gas is necessary to fully reveal the ISM evolution
both in spiral arms and interarm regions \citep{Koda:2011nx, Pety:2013vn}.
The dynamically-driven scenario is built on two observational facts from the
high fidelity imaging.
In interarm regions, molecular emission is abundant compared to atomic emission,
and their ratio shows little gas-phase change between arm and interarm regions.
The MC mass spectrum, however, changes between the two regions,
and the most massive MCs are present only in the spiral arms.
The coagulation and fragmentation of MCs without dissociation of molecules could
naturally explain these two facts \citep{Koda:2009wd}.
More recent CO(1-0) observations of M51's central part \citep{Schinnerer:2013yq}
also confirmed the change of the MC mass spectrum between arm and interarm regions
\citep{Colombo:2014uq}. They re-confirmed that stellar feedback is not the dominant MC
destruction mechanism, since star forming regions are too localized.
Such dynamically-driven ISM evolution is supported by numerical simulations
\citep[e.g., ][]{Kwan:1987qy, Wada:2004fu, Dobbs:2006iq, Wada:2008zh, Tasker:2009dk}.

In this paper we quantify the fraction of molecular gas over total neutral gas (HI+H$_2$)
in the MW and study its azimuthal and radial variations.
The molecular fraction shows a clear radial decrease from the molecule-dominated
center to the atom-dominated outskirts.
The azimuthal variations depend on the globally dominant phase:
in the molecule-dominated inner region ($R\lesssim6\kpc$) the phase changes
only marginally, by only $\sim20\%$, during spiral arm passages.
However, in the atom-dominated outskirts ($R\gtrsim6\kpc$) the phase change
is more pronounced.

\subsection{Brief History}\label{sec:history}
Historically, ISM evolution was often discussed in the context of MC lifetimes and distribution;
it was debated intensely in 1980's
with no consensus emerging \citep[e. g., ][]{Scoville:1979lg, Blitz:1980sh}.
In fact, two major Galactic plane CO(1-0) surveys arrived at contradictory conclusions.
The Columbia/Harvard survey found very little CO emission in the interarm regions
of the longitude-velocity ($l$-$v$) diagram (LVD), which was interpreted as
meaning that MCs survive only for the duration of spiral arm passage
\citep[the order of $\sim 20$-$30$ Myr; ][]{Cohen:1980ve, Blitz:1980sh}.
On the other hand, the Massachusett-Stony Brook survey found
an abundant population of GMCs even in the inter-arm regions, and so concluded
that GMCs live longer than the arm-to-arm transit timescale
\citep[$> 100$ Myr; ][]{Sanders:1985ud, Scoville:2004lr}.

The discussion carries on with more recent $^{13}$CO($J$=1-0) data
\citep{Jackson:2006ya}.
\citet{Roman-Duval:2009fk, Roman-Duval:2010fk} found molecular clouds only on spiral arms
in their face-on map of MW, implying a  short cloud lifetime ($<10$ Myr; their arm crossing time).
Their analysis however suffered from the ambiguity problem of kinematic distance (see Section \ref{sec:circrot}).
Indeed, \citet{Roman-Duval:2010fk} found MC concentrations between the Sagittarius-Carina
and Perseus arms, but considered them as clouds in the Perseus arm due
to possible errors in distance estimation.
Using the same data, \citet{Koda:2006fk} found MCs in the interarm regions using
an analysis of the LVD, which does not involve the distance estimation.
The spiral arm locations in the LVD however are still debatable.

Further progress requires an analysis that does not rely on spiral arm locations
or on distance estimation. In this paper we develop such an analysis.

\subsection{Structures of the Milky Way}

\begin{figure}
\epsscale{1.1}
\plotone{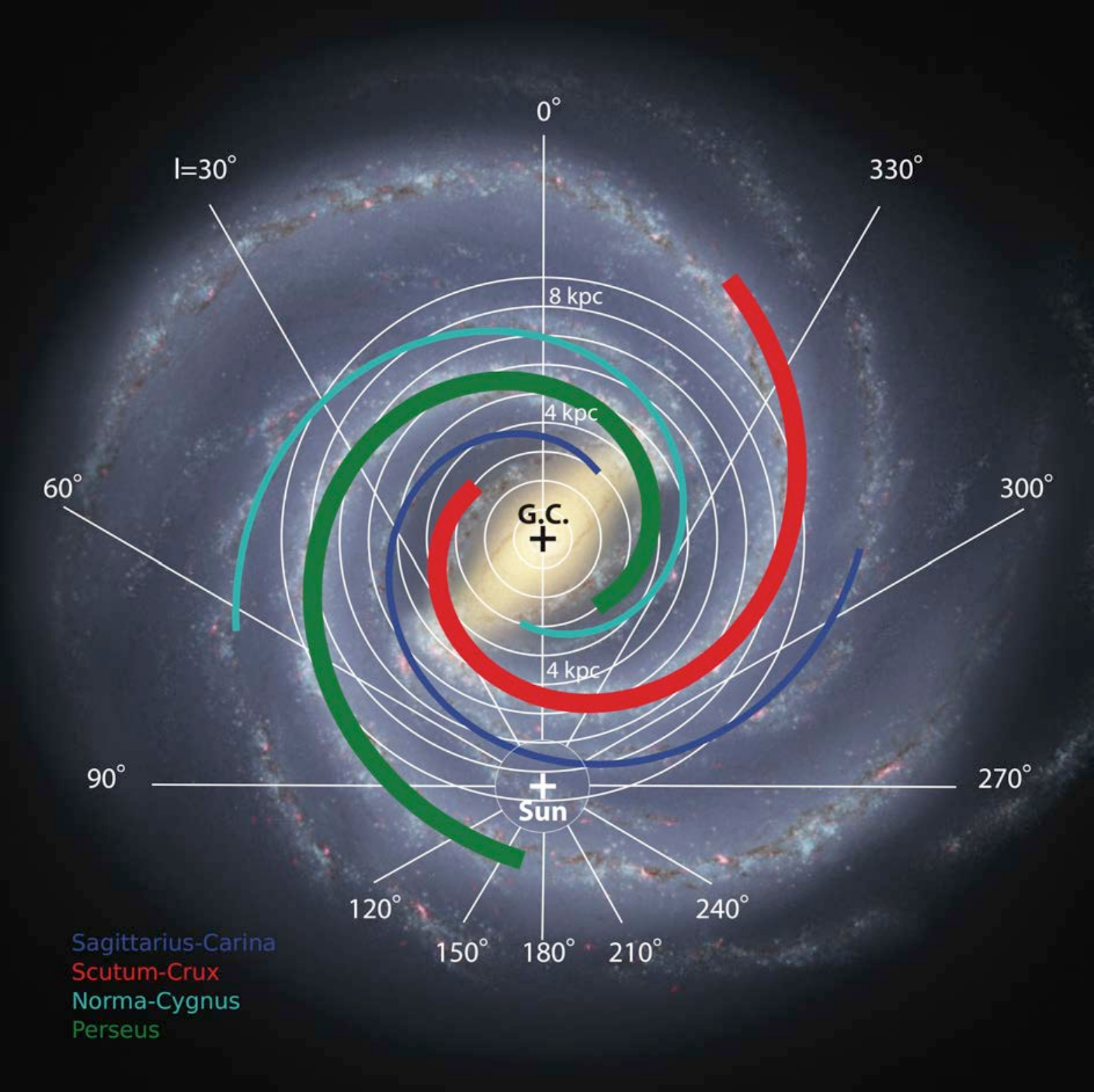}
\caption{Schematic illustration of the face-on view of the Milky Way.
The illustration in the background is from \cite{Churchwell:2009vn}.
The logarithmic spiral arms that we adopted in this paper for qualitative discussions are overplotted.
The blue, red, cyan, and green correspond to the Sagittarius-Carina, Scutum-Crux (aka Scutum-Centaurus),
Norma-Cygnus, and Perseus arms, respectively. Four spiral arms are indicated when tracers of
the interstellar medium (gas or dust) or star formation are used, while only two arms, Scutum-Crux and Perseus,
are traced by the enhancement of stars (see text). Concentric circles around the Galactic Center (G.C.) are
drawn at a 1-kpc interval.
\label{fig:MWspiral_map}}
\end{figure}

The MW is a barred spiral galaxy with a stellar exponential disk scale length of $3.9\pm0.6\kpc$
\citep{Benjamin:2005fr}.
The presence of the central bar was confirmed by stellar distributions.
It has a half-length of $4.4\pm0.5\kpc$ with a tilt of $44\arcdeg \pm10\arcdeg$
to the Sun-Galactic center line \citep{Benjamin:2005fr}.
This corresponds to Galactic longitudes from $l=345\arcdeg$ (or $-15\arcdeg$) to $30\arcdeg$.
Figure \ref{fig:MWspiral_map} shows a conceptual image of the MW in face-on projection
\cite[][]{Churchwell:2009vn} and its spiral arms.

Uncertainties still remain as to the precise number and geometry of spiral arms,
results differing depending on the adopted tracer (c.f. ISM vs. stars).
Four spiral arms are often inferred when gas, dust, or star forming regions
are used as tracers of spiral structure \citep[e.g., ][]{Georgelin:1976mz, Paladini:2004vn, Steiman-Cameron:2010lr, Vallee:2014ve, Hou:2014fj}.
On the other hand, only two arms are indicated by stellar distributions
\citep{Drimmel:2001fk, Benjamin:2008ul, Churchwell:2009vn}.
\citet{Robitaille:2012lr} argued for two major and two minor arms from
a joint-analysis of stellar and gas/dust distributions.
An interpretation of two spiral arms with other two ISM filaments seems
most reasonable, as barred spiral galaxies typically have only two prominent arms.
The ISM in galaxies often shows coherent filamentary structures
even in interarm regions (seen as dust lanes in optical images; e.g., Hubble Heritage images
-- http://heritage.stsci.edu), which explains the minor, apparent gas/dust spiral arms.
In addition, gas filaments in the interarm regions of stellar spiral arms develop
naturally in numerical simulations \citep[e.g., ][]{Kim:2002mb, Chakrabarti:2003mz, Wada:2004fu, Martos:2004aa, Dobbs:2006dg, Pettitt:2015gf}.

Here we adopt the determination of the spiral arms by \citet{Benjamin:2008ul}:
the Perseus and Scutum-Crux (or Scutum-Centaurus) arms are the two major spiral arms
associated with the stellar spiral potential.
The Sagittarius-Carina and Norma-Cygnus features are the two minor spiral arms
without stellar counterparts.
The four arms are drawn in Figure \ref{fig:MWspiral_map}.
We adopt the parameters from \citet{Benjamin:2005fr} and \citet{Churchwell:2009vn},
because of the overall consistency of the bar and spiral arm structures.
We note that there are more recent studies on the MW's structures
\citep[e.g., ][]{Francis:2012aa, Robin:2012aa, Wegg:2015aa}.
For example, in Figure \ref{fig:MWspiral_map},
the inner part of the Perseus arm passes the far end of the bar
and overlaps with the near 3 kpc arm \cite[see ][]{Churchwell:2009vn}.
This section is drawn based on the assumption of symmetry with the Scutum-Crux arm,
but is debatable \citep[e. g. ][]{Vallee:2016aa}.
In any case, our discussion does not depend on the precise definition of the structures.

\section{Data}
In this study, archival HI 21cm and CO($J$=1-0) emission data are used for the analysis of the neutral gas phases.
We focus on the inner part of the MW within the Solar radius:
the range of Galactic longitude $l$ from 0 to 90$\arcdeg$ (the northern part of the inner MW disk)
and from 270 to 360$\arcdeg$ (or equivalently $-90\arcdeg$ to $0\arcdeg$; the southern part).
For most discussions in this paper,
we integrate emission within Galactic latitude $b$ from $-30\arcdeg$ to $+30\arcdeg$,
which covers the full thickness of the gas disk (Section \ref{sec:vertical}).
To determine properties at the Galactic midplane, emission is integrated over $-1.5\arcdeg<b<+1.5\arcdeg$.

The HI 21cm line emission data (3-dimensional data cube) is taken from
the Leiden-Argentine-Bonn (LAB) survey \citep{Kalberla:2005fk}.
The LAB survey is the combination of the Leiden/Dwingeloo survey \citep{Hartmann:1997lr}
and the Instituto Argentino de Radioastronomia Survey \citep{Arnal:2000fk, Bajaja:2005qy}.
The data are corrected for stray radiation picked up by the antenna sidelobes,
and therefore, provides excellent calibration of the 21cm line emission.
This archival data covers the entire sky from $-450$ to $+450\kmps$ at $0.6\arcdeg$ and $1.3\kmps$
resolutions with a spatial pixel size of $0.5\arcdeg$.
From the data, the root-mean-square (RMS) noise on the main beam temperature scale
is estimated to be 0.07-0.09 K.

The CO($J$=1-0) data is from the Columbia/CfA survey \citep{Dame:2001gs}.
This compilation of 37 individual surveys covers all regions with relatively high dust
opacities \citep{Planck-Collaboration:2011qy}, and hence includes virtually all CO
emission in the MW.
The angular resolution and pixel size are $8\arcmin$ ($\sim 0.13\arcdeg$)
and $0.125\arcdeg$, respectively.
The spatial sampling is not uniform and ranges from a full-beam ($0.125\arcdeg$)
in the Galactic plane to super-beam ($0.5\arcdeg$) at higher latitudes.
This does not significantly affect the analyses 
since the CO emission resides predominantly in the thin mid-plane
($|b| \lesssim 5\arcdeg$; see Section \ref{sec:vertical}),
where the sampling is typically $0.125\arcdeg$.
The velocity resolution is $1.3\kmps$.
The RMS noise is 0.12-0.31 K for the inner MW.

\begin{figure}
\epsscale{1.2}
\plotone{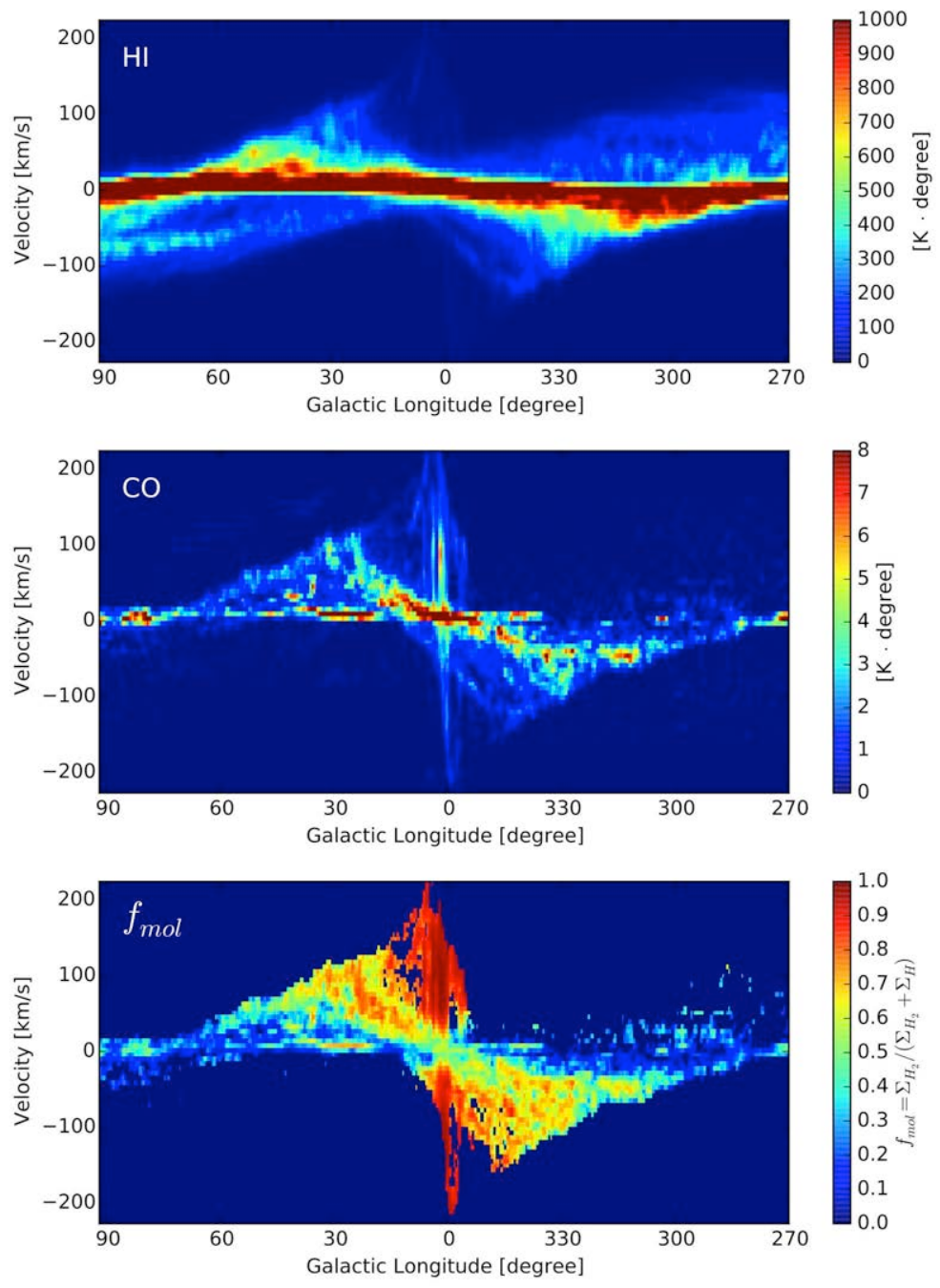}
\caption{Galactic longitude-velocity ($l$-$v$) diagrams (LVDs) of (a) HI 21cm data, (b) CO $J$=1-0 data, and (c) molecular fraction $f_{\rm mol}$.
The $y$-axis is the Local Standard of Rest (LSR) velocity.
The data are smoothed to $0.5\arcdeg$ and $6.2\kmps$ resolutions and integrated over Galactic latitude $-30\arcdeg < b < +30\arcdeg$.
We adopt a CO-to-H$_2$ conversion factor of $X_{\rm CO}=2\times 10^{20} {\rm cm^{-2}\,[K\,\kmps]^{-1}}$.
\label{fig:pv600}}
\end{figure}

\begin{figure*}
\epsscale{1.2}
\plotone{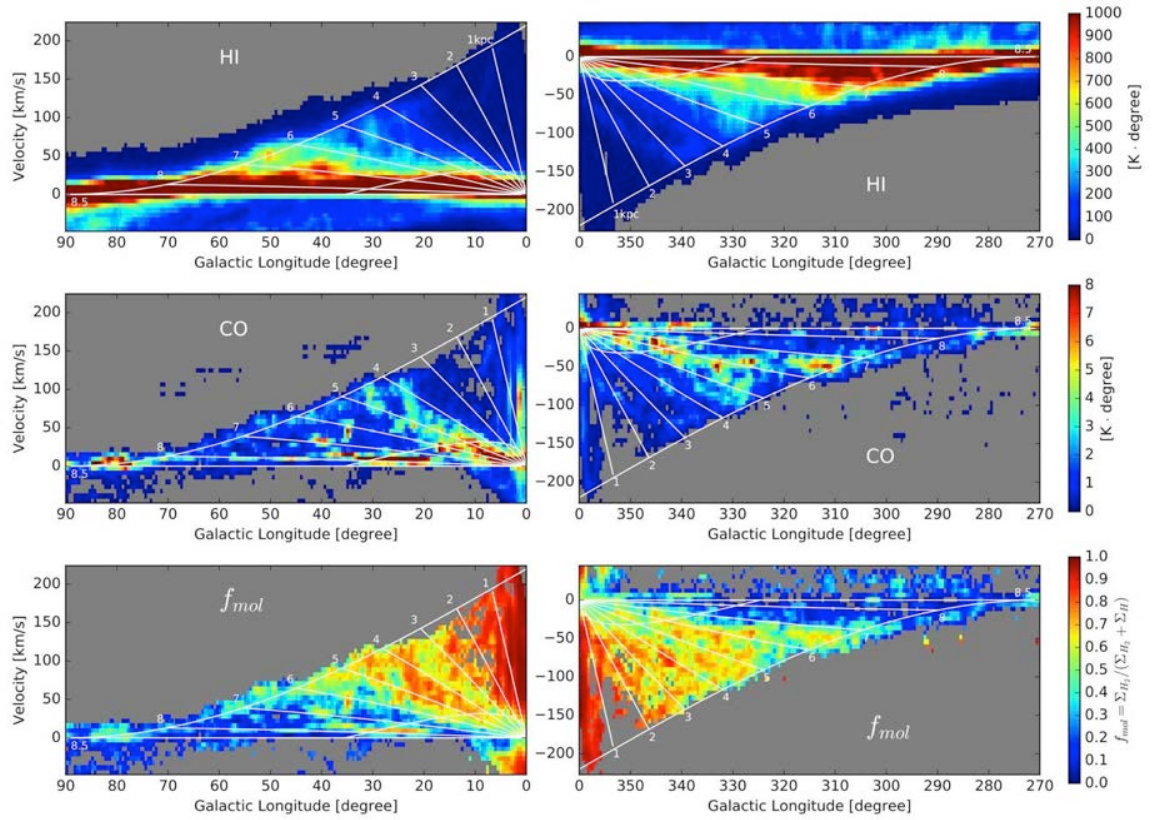}
\caption{The same as Figure \ref{fig:pv600}, but zoomed in on the inner MW.
The northern part (left panels) and southern part of the Galactic plane (right) are separated.
Overplotted are maximum velocity (tangent velocity) curves, as well as Galactocentric
distance lines from $d$=1 to 8 kpc with an 1 kpc interval except the outermost one at 8.5 kpc.
$f_{\rm mol}$ clearly decreases from the Galactic center to the outer part.
The other curves, the arcs from the origins to around $l\sim35\arcdeg$, are the boundaries where
the deviations of $\pm15\kmps$ from a circular rotation result in the errors of 1 kpc in $R$;
we remove the data below these $|v|$ to avoid large errors.
\label{fig:pv600zoom}}
\end{figure*}

To match resolutions, the CO data is smoothed to a $0.6\arcdeg$ resolution
and regridded to a $0.5\arcdeg$ pixel size.
The total integrated flux is conserved in these operations.
Both HI and CO cubes are then binned at a $6.2 \kmps$ velocity resolution.
The width of the velocity bin is chosen so that local fluctuations,
e.g., due to molecular clouds with a typical velocity dispersion of $\sim 4\kmps$ \citep[ for giant molecular clouds; ][]{Scoville:1987vo},
are smoothed out, but the spiral arms and interarms are still resolved (see Section \ref{sec:noncirc} and Appendix).
The spatial resolution of $0.6\arcdeg$ corresponds to a linear scale of
$104 (d/10\kpc) \pc$ for the heliocentric distance $d$ in kpc.
Note that  we integrate the data over a larger range in Galactic latitude ($b$).
Figure \ref{fig:pv600}a,b show the LVDs of HI and CO emission.
The smoothed data cubes are integrated over 
$-30\arcdeg < b < +30\arcdeg$ for these figures.
Figure \ref{fig:pv600zoom} is the same as Figure \ref{fig:pv600}, but shows
a zoom-in on the inner MW.

\section{Method}\label{sec:method}

Our goal is to analyze variations of the molecular gas fraction $f_{\rm mol}$ (defined in Section \ref{sec:deffmol})
in the radial and azimuthal directions.
In this section, we first calculate $f_{\rm mol}$ in an LVD (Section \ref{sec:deffmol}; Figure \ref{fig:pv600zoom}c),
mapping each ($l$, $v$) pixel into a Galactcentric radius $R$ on the assumption of Galactic circular rotation (Section \ref{sec:circrot}),
and then producing an $f_{\rm mol}$-$R$ plot (Figure \ref{fig:radial}).

This direct conversion of $f_{\rm mol}$ in an LVD to that in $R$ has clear advantages.
First, it does not require heliocentric distances, since $f_{\rm mol}$ is a distance-independent parameter.
A more conventional analysis of MW structures resolves the near-far distance ambiguity in kinematic
distance measurement.
While this offers a measure of CO or HI luminosities, and therefore, mass, it can also generate
systematic uncertainties.
Second, the LVD samples both spiral arms and interarm regions (Section \ref{sec:noncirc})
even though their exact locations are uncertain.
The $f_{\rm mol}$-$R$ plot (Figure \ref{fig:radial}) should therefore include data from both
arm and interarm regions. Therefore, variations of $f_{\rm mol}$ at a fixed radius in this plot
correspond to azimuthal variations, i.e., arm/interarm variations, at that radius.
This way, the radial and azimuthal variations of $f_{\rm mol}$ are quantified
without identifying the exact locations of spiral arms and interarm regions.

Potential errors in our analysis are sumarized in Section \ref{sec:syserror}, and here
we note one caveat.
Each velocity at a given $l$ corresponds to two locations along the line-of-sight
(i.e., near and far sides; see Section \ref{sec:circrot}).
They are averaged in the $f_{\rm mol}$ calculation.
This is not a problem when both near and far sides correspond to either arm or interarm regions.
An LVD shows two types of regions (see Section \ref{sec:noncirc}): one in which
both near and far sides are interarm regions, and the other where
a spiral arm and interarm region overlap.
For the latter case, spiral arm emission is most likely dominant,
and their average should represent $f_{\rm mol}$ of the spiral arm.

\subsection{The Molecular Fraction}\label{sec:deffmol}

The azimuthally-averaged radial trend of gas phase in galaxies has been studied 
with two different expressions of the fraction of molecular gas
\citep{Elmegreen:1993fk, Sofue:1995fk, Wong:2002lr, Blitz:2006or}.
We use the definition adopted by \citet{Sofue:1995fk}.
The molecular fraction $f_{\rm mol}$, i.e., the mass fraction of H$_2$ gas over the total HI+H$_2$ gas, is
expressed as
\begin{equation}
f_{\rm mol} \equiv \frac{\Sigma_{\rm H_2}}{\Sigma_{\rm H_2}+\Sigma_{\rm HI}},\label{eq:fmol}
\end{equation}
where $\Sigma_{\rm H_2}$ and $\Sigma_{\rm HI}$ are the surface densities
of H$_2$ and HI gas, respectively.
The $\Sigma_{\rm H_2}$ and $\Sigma_{\rm HI}$ [$\Msun \,\rm pc^{-2}$] are calculated from the HI and CO integrated intensities,
$I_{\rm HI}$ and $I_{\rm CO}$ [K km/s], at each ($l$, $v$) pixel, respectively.
On the assumption of optically-thin HI 21cm emission,
\begin{equation}
\Sigma_{\rm HI} = 1.45\times 10^{-2} I_{\rm HI} \left(\frac{dv}{dd}\right),\label{eq:sighi}
\end{equation}
where the expression in the parenthesis is a derivative of $v$ with respect to $d$.
Using the CO-to-H$_2$ conversion factor $X_{\rm CO}$,
we have
\begin{equation}
\Sigma_{\rm H_2} = 3.2 I_{\rm CO} \left( \frac{X_{\rm CO}}{2\times 10^{20} \rm \,H_2 \,cm^{-2} \, [K\cdot \kmps]^{-1}} \right)  \left(\frac{dv}{dd}\right). \label{eq:sigco}
\end{equation}
We use $X_{\rm CO}=2\times 10^{20} \rm \,H_2 \,cm^{-2} \, [K\cdot \kmps]^{-1}$
as recommended by \citet{Bolatto:2013ys}; $f_{\rm mol}$ increases by $\sim 10\%$ at some radii
if an $X_{\rm CO}$ of $3\times 10^{20} \rm \,H_2 \,cm^{-2} \, [K\cdot \kmps]^{-1}$ is adopted instead (see Section \ref{sec:caveats}).
The "$dv/dd$" term is the line-of-sight velocity gradient and converts $I_{\rm HI}$ and $I_{\rm CO}$
in a velocity bin to the values in a face-on projection of the MW disk  \citep[e.g., ][]{Nakanishi:2003xw, Nakanishi:2006uq}.
Our analysis does not suffer from the systematic uncertainty due to this term
since it cancels out in eq. (\ref{eq:fmol}).
We do not include He mass in eq. (\ref{eq:sighi} and \ref{eq:sigco}) since it also cancels out in eq.  (\ref{eq:fmol}).

Figure \ref{fig:pv600}c and \ref{fig:pv600zoom}c show $f_{\rm mol}$ in LVDs.
The sensitivities in HI and CO change within the diagrams.
Our analysis is limited by the sensitivity of the CO observations.
We derived a conservative RMS noise estimate roughly from emission-free pixels
and cut the pixels below $\sim$3 $\sigma$ significance, i.e.,
$0.17 {\rm \,K\cdot degree}$ for CO.
The sensitivity should be higher in the inner part that we analyze in this study.
The HI emission is detected much more significantly at all locations,
and for Figure \ref{fig:pv600} we imposed the cut-off at a threshold surface density
$\sim 7$ times lower than that for CO.
The pixels with the low signal-to-noise are blanked in these figures
and not included in the subsequent analysis.
This excludes virtually no point between the Galactocentric radii
$3.5\lesssim R \lesssim 8.5 \kpc$, but does some at $R\lesssim 3.5\kpc$ (see Section \ref{sec:circrot}).
The removed points correspond to the radii inside those of the Galactic bar, but azimuthally not
in the bar. Gas is often deficit there in other barred galaxies \citep[e.g., ][]{Sheth:2002lr}.

\subsubsection{Notes on Dark Gas and $X_{\rm CO}$}\label{sec:caveats}

The presence of dark gas, invisible in CO or HI emission, is inferred
from $\gamma$-ray, submm surveys of dust emission, and far-IR spectroscopy
of [CII] emission. \citep{Grenier:2005lr, Planck-Collaboration:2011qy, Pineda:2013lr}.
Its mass fraction is estimated to be about 22\% in the solar neighborhood.
This gas could be the CO-deficient H$_2$ gas \citep{van-Dishoeck:1988br, Wolfire:2010fk}
or the optically-thick HI gas \citep{Fukui:2014wd, Fukui:2015fk}, and the reality is perhaps a mixture
of both. Our analysis does not include the dark gas,
but the dark H$_2$ and HI gas should compensate each other in the $f_{\rm mol}$
calculation if their fractions are similar.

The actual $X_{\rm CO}$ in the MW disk could be larger than the one recommended by \citet{Bolatto:2013ys}.
Among all measurements, only two, those based on virialized MCs and $\gamma$-ray emission,
attempt to anchor their calibrations with actual HI+H$_2$ mass measurements.
The others rely on scalings with abundances or gas-to-dust ratios without measuring the mass.
Those scaling constants are at least as uncertain as $X_{\rm CO}$.
The average value among virialized MCs is $X_{\rm CO} =3.6\times 10^{20}$
\citep{Scoville:1987lp} or $3.5\times 10^{20}$ \citep[][ using $R_0=8.5\kpc$]{Solomon:1987pr},
instead of the value that \citet{Bolatto:2013ys} derived for a typical GMC from the same data
(we omit the unit, $\rm H_2 \,cm^{-2} \, [K\cdot \kmps]^{-1}$).
The background cosmic ray distribution for $\gamma$-ray production is still uncertain,
and $X_{\rm CO}$ from $\gamma$-ray suffers from this uncertainty.
There is also a possibility that $X_{\rm CO}$ is smaller in the central molecular zone
of the Milky Way.
The following equation converts $f_{\rm mol}$ with $X_{\rm CO}=2\times 10^{20}$
to the value $f_{\rm mol}^{\prime}$ with some other $X_{\rm CO}^{\prime}$,
\begin{equation}
f_{\rm mol}^{\prime} = \frac{(X_{\rm CO}^{\prime}/X_{\rm CO})f_{\rm mol}} {(X_{\rm CO}^{\prime}/X_{\rm CO})f_{\rm mol} + (1-f_{\rm mol})}.
\end{equation}
Table \ref{tab:fmol} presents $f_{\rm mol}$ using the consensus $X_{\rm CO}$ value and a slightly larger one (see Section \ref{sec:radial}).
The differences do not affect the main conclusions of this paper.

\subsection{The Constant Circular Rotation Model}\label{sec:circrot}

\begin{figure}
\epsscale{1.2}
\plotone{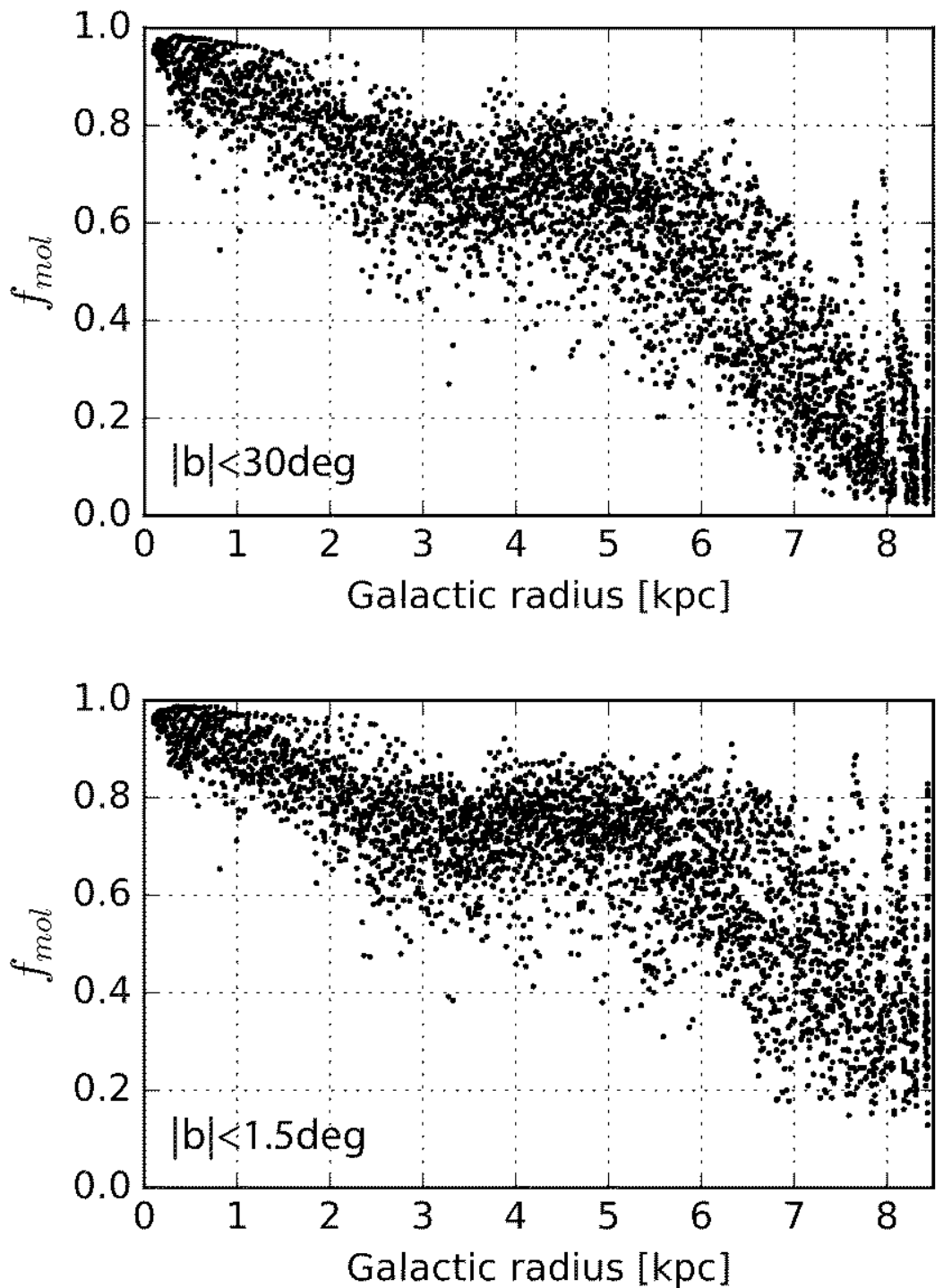}
\caption{The molecular fraction $f_{\rm mol}$ as a function of $R$.
The atomic and molecular gas is integrated
(a) over the whole disk thickness $|b|<30\arcdeg$ and
(b) around the midplane $|b|<1.5\arcdeg$.
The coordinates ($l$, $v$) of each pixel in the LVD (Figure \ref{fig:pv600zoom}) are
converted to $R$ using eq. (\ref{eq:rotv}).
The radially-decreasing trend is clear in both panels.
The scatter at each radius represents variations of $f_{\rm mol}$ along the
circular orbit at that Galactocentric radius, therefore showing the amount of azimuthal variations,
and hence arm/interarm variations. 
\label{fig:radial}}
\end{figure}

Under the assumption of constant circular rotation of the MW,
a simple geometric consideration provides the equation
for conversion from ($l$, $v$) to $R$  \citep[e.g., ][]{Oort:1958fj, Kellermann:1988lr, Nakanishi:2003xw},
\begin{equation}
R = R_0 \frac{V_0 \sin l}{v + V_0 \sin l},
\label{eq:rotv}
\end{equation}
where we set the constant rotation velocity $V_0=220\kmps$ and
the Sun's location at $R_0=8.5\kpc$.
This equation is applicable in a particular velocity range at each $l$ (explained later):
in summary,
$0\leq v \leq V_0 ( 1- \sin l)$ for $0 \leq l \leq 90\arcdeg$
and 
$-V_0 ( 1+ \sin l) \leq v \leq 0$ for $270 \leq l \leq 360\arcdeg$.
Figure \ref{fig:pv600zoom} shows LVDs with lines of constant $R$ at 1 kpc intervals
and at $R=8.5\kpc$.
Figure \ref{fig:radial}(a) shows the $f_{\rm mol}$-$R$ plot from Figure \ref{fig:pv600zoom}c,
integrated over the whole gas disk thickness ($|b|<30\arcdeg$; Section \ref{sec:vertical}),
and Figure \ref{fig:radial}(b) is for the Galactic midplane $|b|<1.5\arcdeg$.

We do not use the heliocentric distance $d$ except for the calculation of vertical profiles
of the molecular and atomic gas at tangent points (defined below).
$d$ and $R$ are related by
\begin{equation}
d = R_0 \cos l \pm \sqrt{R^2 - R_0^2\sin^2l}.
\label{eq:dist}
\end{equation}
for the inner MW (i.e., $R<R_0$).
Two solutions,  "$\pm$" for near and far distances, 
are possible for a given $R$, hence for a given ($l$,$v$); this is the near-far distance ambiguity.

\begin{figure*}
\epsscale{1.2}
\plotone{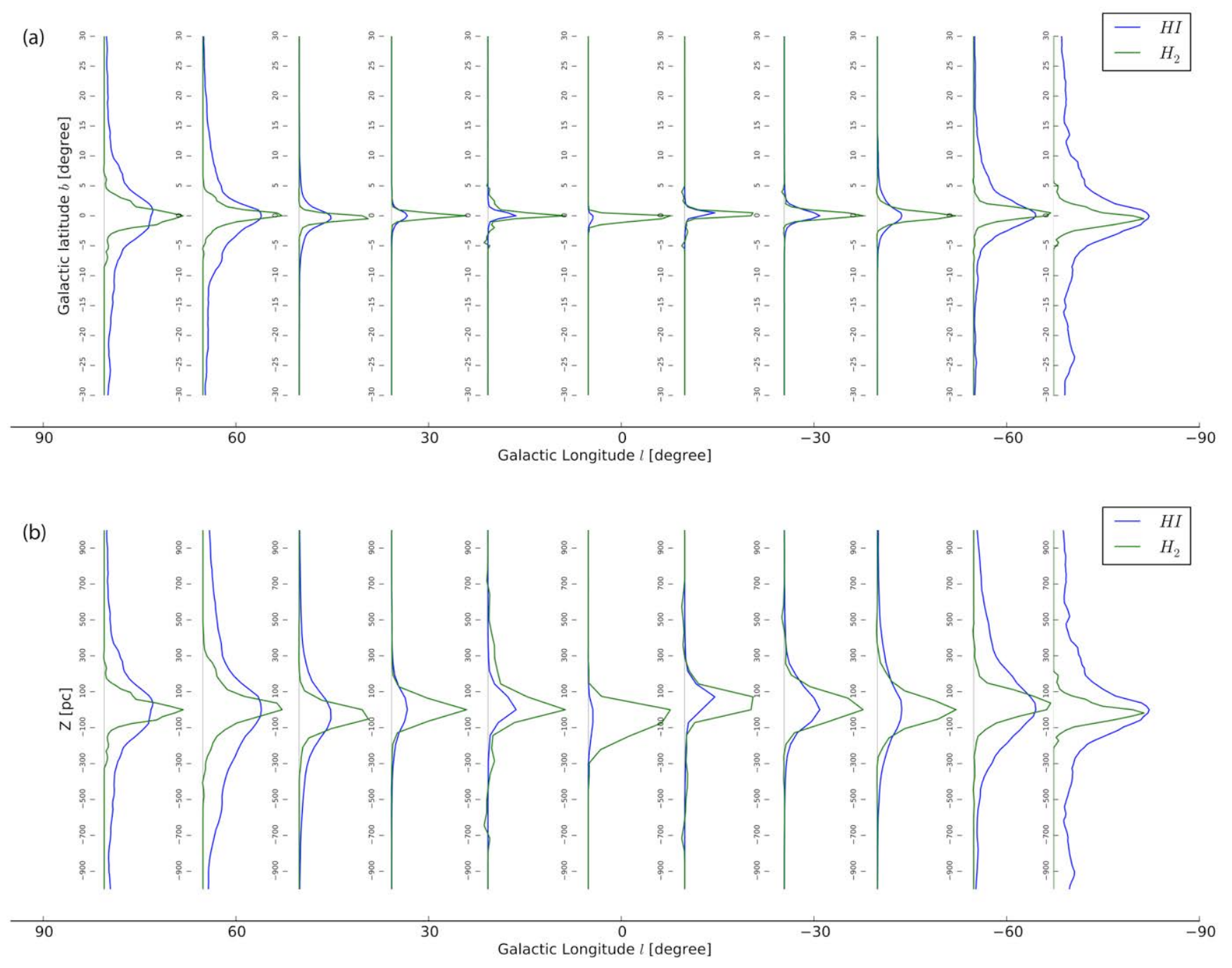}
\caption{Vertical profiles of atomic (HI -blue) and molecular (H$_2$ -green) gas surface densities (eqs. \ref{eq:sighi} and \ref{eq:sigco})
at tangent points (eq. \ref{eq:tangent})
at $|l|=$0, 15, 30, 45, 60, and 75 $\arcdeg$, corresponding to Galactocentric distances of $R=$0, 2.2, 4.2, 6.0, 7.4, and 8.2$\kpc$
and heliocentric distances of $d=$8.5, 8.2, 7.4, 6.0, 4.3, and 2.2$\kpc$.
Each vertical profile is an average over $\pm5\arcdeg$ in $l$ and $\pm20\kmps$ in $v$ around a tangent point.
The absolute amplitude scale of each plot is scaled arbitrarily, though the relative amplitudes between HI and H$_2$ are preserved
within each plot.
The vertical axis is (a) in degrees, covering $-30\arcdeg<b<30\arcdeg$, and (b) in parsecs, covering -1000 pc to 1000 pc in the direction
of the disk thickness.
\label{fig:vertical}}
\end{figure*}

The line-of-sight at $l$ is tangential to the circular orbit of radius $R= R_0 |\sin l |$.
Their intersection is called a tangent point.
The line-of-sight velocity $v$ in the direction of $l$ takes its maximum absolute value
at the tangent point,
\begin{equation}
v=
\left\{
\begin{array}{ll}
V_0 ( 1- \sin l) & (0 \leq l \leq 90\arcdeg) \\
-V_0 ( 1+ \sin l) & (270 \leq l \leq 360\arcdeg),
\end{array}
\right.
\label{eq:tangent}
\end{equation}
which set the velocity range for eq. (\ref{eq:rotv}).
These maximum line-of-sight velocities (i.e., tangent velocities) are drawn in Figure \ref{fig:pv600zoom}.
There is no distance ambiguity problem at tangent points  ($d = R_0 \cos l$ from eq. \ref{eq:dist}).
Figure \ref{fig:vertical}b shows the vertical profiles
of HI and H$_2$ gas masses at the tangent points, which is discussed in Section \ref{sec:vertical}.

\subsubsection{Removal of Regions of Potentially Large Errors}

The constant $R$ lines are crowded around small $|l|$ and $|v|$ in Figure \ref{fig:pv600zoom}.
We remove these regions since a small deviation from circular rotation would result in a large error in $R$.
By taking the derivative of eq. (\ref{eq:rotv}) we derive the relation between a slight shift $\Delta v$
and resultant error $\Delta R$. We then obtain the equation for removal,
\begin{equation}
\left| \frac{\Delta R}{\Delta v} \right| = \left| \frac{R_0V_0 \sin l}{ (v+V_0 \sin l)^2} \right| > \gamma
\label{eq:removal}
\end{equation}
Arbitrarily we set $\gamma = 1/15\, [\kpc\,(\kmps)^{-1}$], meaning that an ($l$,$v$)
pixel is removed if a $15\kmps$ deviation in $v$ from the circular rotation causes an error larger than 1 kpc in $R$.
These threshold lines are drawn in Figure \ref{fig:pv600zoom}, as arcs that start from the origin
and run to about $l\sim \pm35\arcdeg$; the pixels with smaller $|l|$ and $|v|$ are removed.

\subsection{Spiral Arms and Non-Circular Motions}\label{sec:noncirc}
Gas motions are approximated with circular rotation in our analysis,
and we do not include non-circular motions associated with the MW spiral arms.
The spiral arms and associated non-circular motions
affect the LVD in a systematic way.
Their impact is small as demonstrated in Appendix \ref{sec:modelvld}.
In this model, two points are important to keep in mind.

First, the locations of spiral arms systematically shift in the velocity direction
on an LVD due to non-circular motions
(the ``loops" of spiral arms in LVD are squashed in the velocity direction).
This does not change the areas that the spiral arms occupy in an LVD, but
causes errors in $R$ if it is determined in assumption of a circular rotation.
This results in an artificial steepening of the radial gradient of $f_{\rm mol}$
near the tangent radii of the spiral arms.
[See Figure \ref{fig:pvmodel}c,d in Appendix --
if the black solid lines represent the intrinsic radial profile of $f_{\rm mol}$,
the error vectors show how the points on the profile apparently shift
due to the non-circular motions.
By connecting the vector heads, we see an apparently-steepened radial profile.
The vector directions change from the right to left near the tangents of the spiral arms.]

\begin{figure}
\epsscale{1.1}
\plotone{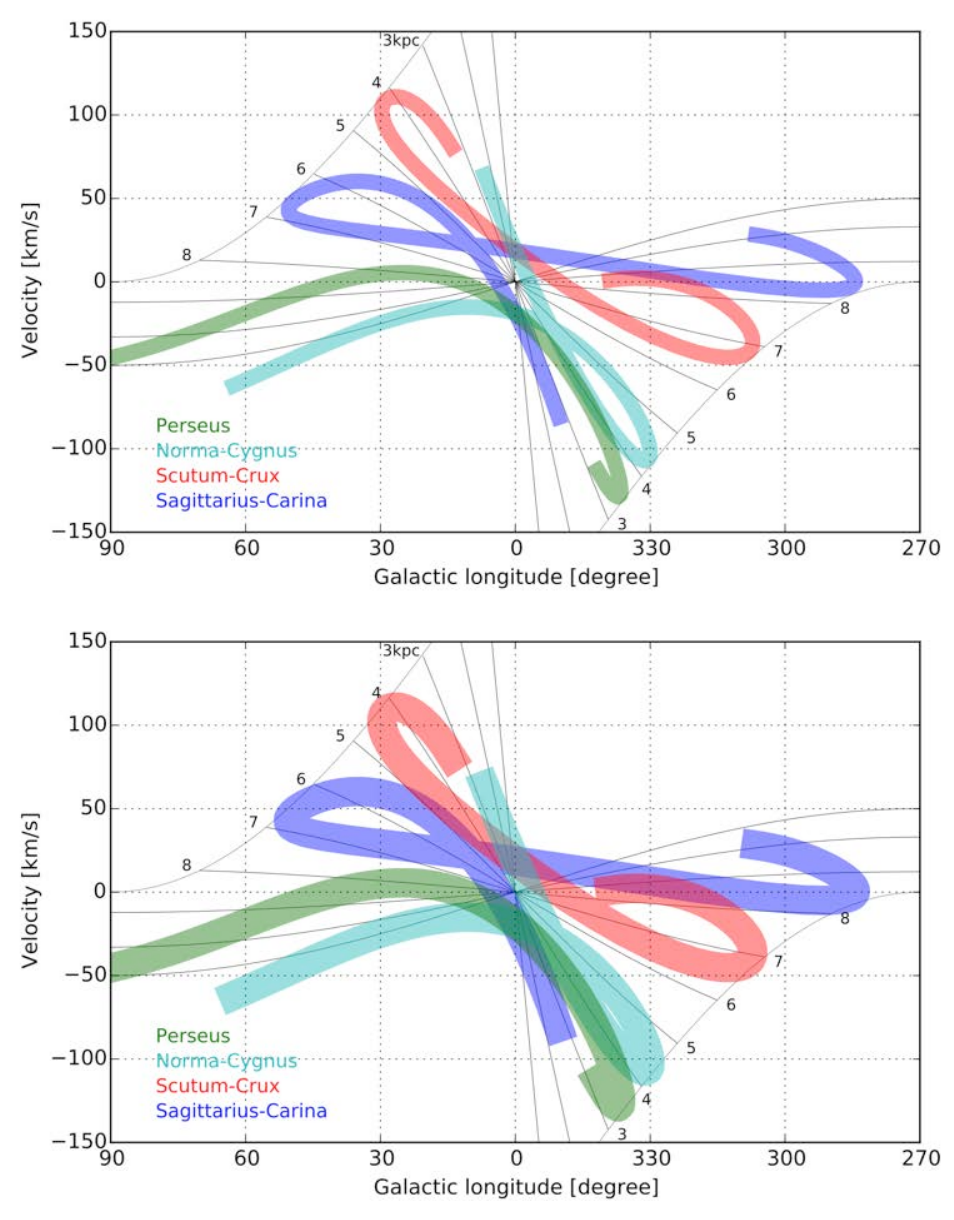}
\caption{Model spiral arms on an LVD. Non-circular motions and velocity widths (due to
increased velocity dispersions) in spiral arms are included as discussed in Section \ref{sec:noncirc} and Appendix \ref{sec:modelvld}.
Only the Perseus and Scutum-Crux arms have corresponding stellar spiral arms.
The Norma-Cygnus and Sagittarius-Carina arms are seen only in the ISM distribution, but not as
stellar spiral arms, and might be interarm gas structures between the two stellar spiral arms.
Spiral arms with
(a) velocity widths that correspond roughly to the observed ones: about $\sim10\kmps$
when the arm are seen perpendicular to our line of sight and
$\sim20$-$30\kmps$ around their tangent positions,
(b) the maximum possible velocity widths: $\sim 20\kmps$ when perpendicular to
the line of sight, and up to $\sim 50\kmps$ at the tangent positions.
\label{fig:pvmodel_width}}
\end{figure}

Second, even with increased velocity widths due to enhanced velocity dispersions and spiral arm
streaming, interarm regions are still sampled in the LVD (Figure \ref{fig:pvmodel_width}),
Thus, the $f_{\rm mol}$-$R$ plot from the LVD includes the data both from spiral arms and interarm regions.

\section{Results}\label{sec:results}

\subsection{Vertical Profiles}\label{sec:vertical}

Figure \ref{fig:vertical} shows the vertical profiles of atomic (HI) and molecular (H$_2$) gas masses at tangent points
at a 15$\arcdeg$ interval in $l$. The amplitude scales for the HI and H$_2$ profiles are the same
in mass density within each panel; but different scalings are used in the different panels.
Regions of $\pm5\arcdeg$ in $l$ and $\pm 20\kmps$ in $v$ are integrated at the tangent points (i.e., at the terminal velocities).
Note that from low to high $l$ the Galactocentric distance $R$ increases.
Panel (a) shows the distributions as a function of Galactic latitude $b$.
Both HI and CO are confined within $-30\arcdeg < b < 30\arcdeg$, over which we integrate HI and CO
emission for the LVDs (except for Figure \ref{fig:radial}b where only the midplane $-1.5\arcdeg < b < 1.5\arcdeg$
is integrated).

In a parsec scale, the molecular gas is confined in the thin mid-plane at all $l$ (and $R$), while the atomic gas
is distributed over a thicker disk.
Figure \ref{fig:vertical}(b) shows the vertical distributions with the Galactic altitude $Z$ from the mid-plane in parsecs.
The FWHM thicknesses of the atomic and molecular disks within the Solar radius ($R=8.5\kpc$)
were measured using functional fits (e.g., Gaussian or sech$^2$) and are $\sim 100$-$200\pc$ and $\sim 50$-$100\pc$, respectively
\citep{Sanders:1984fx, Dickey:1990fk, Nakanishi:2003xw, Nakanishi:2006uq, Kalberla:2007aa}.
This is consistent with the results in Figure \ref{fig:vertical}b given that our $0.6\arcdeg$ resolution corresponds to
$\sim 90 \pc$ at the Galactic center distance ($d=8.5\kpc$).

The molecular gas is always the major phase ($\gtrsim 50\%$ in mass) at the mid-plane, from the center to the Solar radius.
The mid-plane of the gas disk moves slightly up and down locally \citep{Sanders:1984fx, Nakanishi:2003xw, Nakanishi:2006uq}.
By adopting the locations of the profile peaks as the approximate midplane,
the H$_2$ mass always dominates the HI mass in the inner MW; starting from $f_{\rm mol}\sim 100\%$ in the central region
and becoming comparable to HI, $f_{\rm mol}\sim 50\%$, around $|l|=60$-$75\arcdeg$ ($R=7.4$-$8.2\kpc$).
Figure \ref{fig:radial}b also shows the dominance of the molecular gas in the mid-plane from the center to the solar radius.
The dominance of H$_2$ ends around $R\gtrsim6\kpc$ when the gas at high altitudes is included (see Section \ref{sec:radial}),
since the HI gas becomes abundant at high altitudes.

\subsection{Radial Profile}\label{sec:radial}

\begin{figure}
\epsscale{1.2}
\plotone{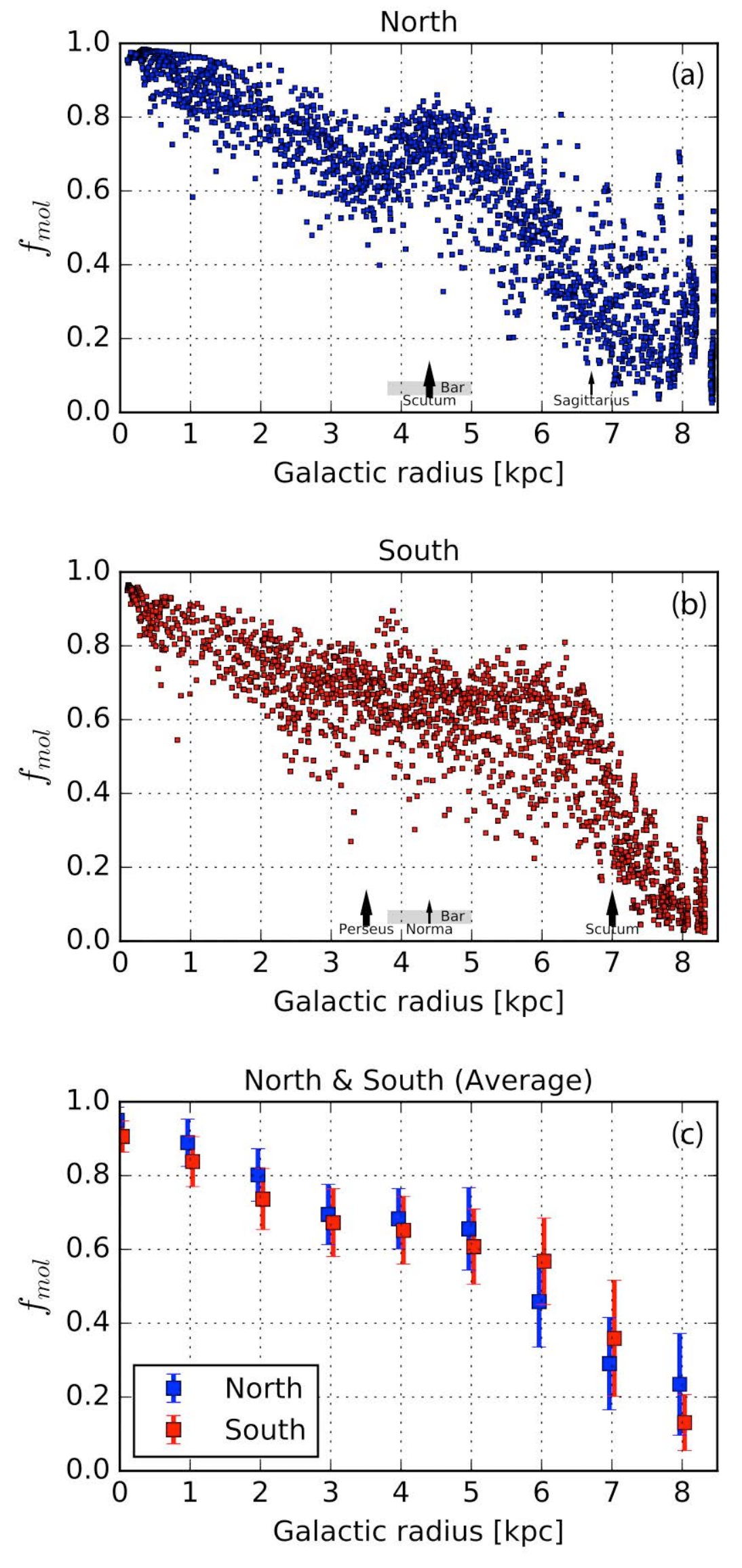}
\caption{The same as Figure \ref{fig:radial}a (integrated over the whole disk thickness $|b|<30\arcdeg$), but
(a) for the northern part of the Galactic plane,
(b) for the southern part, and
(c) showing averages and standard deviations in 1 kpc bins for the northern and southern parts
[The points are slightly shifted in the horizontal direction to reduce overlaps].
The tangent points of the four spiral arms are indicated by arrows;  the thick arrows indicate
the two arms identified as stellar enhancements (i.e., potential).
The bar end, $4.4\pm0.5\kpc$, is indicated with a gray horizontal line.
Only the tangent points of the spiral arms are indicated, but the data from other parts of the arms
are distributed over all radii.
\label{fig:radialns}}
\end{figure}

The Galactocentric distance appears to be the most important parameter for variations of $f_{\rm mol}$.
Figure \ref{fig:radial} show the $f_{\rm mol}$-$R$ plots with the data integrated over (a) the whole disk thickness $|b|<30\arcdeg$
and (b) around the midplane $|b|<1.5\arcdeg$. In both cases, a radially-decreasing trend is very clear.
Figure \ref{fig:radialns} displays the northern and southern sides of the inner MW disk separately
for $|b|<30\arcdeg$ and confirms that the radial dependence is the determinant.
All HI and H$_2$ gas within $|b|<30\arcdeg$ is integrated for these plots.
The radial trend is also seen in the LVD (Figure \ref{fig:pv600zoom}c);
declining from the innermost wedge at $R=0$-$1\kpc$ to the outermost one.

The radial decrease and consistency between the northern and southern disks
are clearer in Figure \ref{fig:radialns}(c).
It shows the average and RMS scatter in radial bins with a 1-kpc width, except for the central bin with a 0.5-kpc width.
Quantitatively, $f_{\rm mol}$ decreases from about $\sim 100\%$ at the center,
remaining $\gtrsim 50\%$ out to $R\sim6\kpc$, and decreasing to $\sim 10$-$20\%$ at $R\sim 8.5\kpc$,
when integrated over the whole disk thickness $|b|<30\arcdeg$.
The largest difference between the two sides is only about 11\% at $R=6\kpc$,
and hence azimuthal variations are small.
Table \ref{tab:fmol} lists $f_{\rm mol}$ as a function of the Galactocentric radius $R$.

Previous studies analyzed the molecular fraction in galaxies on an azimuthally-averaged basis
and found similar radial trends \citep{Sanders:1985ud, Young:1991kq, Sofue:1995fk, Honma:1995qy, Wong:2002lr}.
The dominance of the radial dependence over an azimuthal one in the MW
became clear in Figure \ref{fig:radial}.

\begin{deluxetable*}{cccccccc}
\tablecolumns{8}
\tablewidth{0pc}
\tablecaption{Molecular Fraction $f_{\rm mol}$}
\tablehead{
\colhead{$R$} & \multicolumn{3}{c}{Midplane}  & \colhead{}& \multicolumn{3}{c}{Whole Disk Thickness}  \\
\cline{2-4} \cline{6-8}\\
\colhead{(kpc)} & \colhead{North}   & \colhead{South}  & \colhead{Average} & \colhead{}  & \colhead{North} & \colhead{South}  & \colhead{Average}}
\startdata
&  \multicolumn{7}{c}{$X_{\rm CO}=2\times 10^{20} {\rm cm^{-2}\,[K\,\kmps]^{-1}}$}   \\
\cline{2-8} \\
0.0 & $0.96\pm0.03$ & $0.92\pm0.03$ & $0.94\pm0.04$ && $0.95\pm0.04$ & $0.91\pm0.04$ & $0.93\pm0.04$ \\
1.0 & $0.91\pm0.05$ & $0.88\pm0.06$ & $0.90\pm0.06$ && $0.89\pm0.06$ & $0.84\pm0.07$ & $0.87\pm0.07$ \\
2.0 & $0.84\pm0.07$ & $0.77\pm0.08$ & $0.81\pm0.08$ && $0.80\pm0.07$ & $0.74\pm0.08$ & $0.77\pm0.08$ \\
3.0 & $0.74\pm0.08$ & $0.71\pm0.09$ & $0.72\pm0.08$ && $0.69\pm0.08$ & $0.67\pm0.09$ & $0.68\pm0.09$ \\
4.0 & $0.74\pm0.08$ & $0.71\pm0.08$ & $0.73\pm0.08$ && $0.68\pm0.08$ & $0.65\pm0.09$ & $0.67\pm0.09$ \\
5.0 & $0.74\pm0.10$ & $0.70\pm0.10$ & $0.72\pm0.10$ && $0.66\pm0.11$ & $0.61\pm0.10$ & $0.63\pm0.11$ \\
6.0 & $0.61\pm0.10$ & $0.70\pm0.11$ & $0.65\pm0.11$ && $0.46\pm0.12$ & $0.57\pm0.12$ & $0.51\pm0.13$ \\
7.0 & $0.49\pm0.14$ & $0.54\pm0.17$ & $0.52\pm0.16$ && $0.29\pm0.12$ & $0.36\pm0.16$ & $0.32\pm0.15$ \\
8.0 & $0.46\pm0.17$ & $0.37\pm0.13$ & $0.43\pm0.16$ && $0.23\pm0.14$ & $0.13\pm0.07$ & $0.19\pm0.13$ \\
\cline{2-8} \\
&  \multicolumn{7}{c}{$X_{\rm CO}=3\times 10^{20} {\rm cm^{-2}\,[K\,\kmps]^{-1}}$}   \\
\cline{2-8} \\
0.0 & $0.97\pm0.02$ & $0.95\pm0.02$ & $0.96\pm0.02$ && $0.97\pm0.02$ & $0.94\pm0.03$ & $0.95\pm0.03$ \\
1.0 & $0.94\pm0.04$ & $0.92\pm0.04$ & $0.93\pm0.04$ && $0.92\pm0.05$ & $0.89\pm0.05$ & $0.91\pm0.05$ \\
2.0 & $0.89\pm0.05$ & $0.84\pm0.06$ & $0.86\pm0.06$ && $0.86\pm0.05$ & $0.81\pm0.07$ & $0.83\pm0.07$ \\
3.0 & $0.81\pm0.06$ & $0.79\pm0.07$ & $0.80\pm0.07$ && $0.77\pm0.07$ & $0.75\pm0.08$ & $0.76\pm0.07$ \\
4.0 & $0.81\pm0.06$ & $0.79\pm0.07$ & $0.80\pm0.07$ && $0.76\pm0.07$ & $0.74\pm0.08$ & $0.75\pm0.07$ \\
5.0 & $0.81\pm0.08$ & $0.78\pm0.08$ & $0.79\pm0.08$ && $0.74\pm0.09$ & $0.70\pm0.09$ & $0.72\pm0.09$ \\
6.0 & $0.70\pm0.09$ & $0.77\pm0.09$ & $0.74\pm0.10$ && $0.56\pm0.12$ & $0.66\pm0.11$ & $0.61\pm0.12$ \\
7.0 & $0.59\pm0.13$ & $0.64\pm0.16$ & $0.62\pm0.15$ && $0.38\pm0.14$ & $0.46\pm0.17$ & $0.42\pm0.16$ \\
8.0 & $0.56\pm0.17$ & $0.47\pm0.14$ & $0.53\pm0.17$ && $0.31\pm0.17$ & $0.18\pm0.10$ & $0.26\pm0.16$
\enddata
\label{tab:fmol}
\end{deluxetable*}

The MW has a bar structure with a half-length of $4.4\pm0.5\kpc$ \citep{Benjamin:2005fr}.
Barred spiral galaxies often show bright CO emission along their bars and at the bar ends \citep[e.g., ][]{Sheth:2002lr}.
On the northern side (Figure \ref{fig:radialns}a), the most prominent bump
around $R\sim 4.5\kpc$, $\sim 20\%$ enhancement, may correspond the bar end.
A similar enhancement is seen in the southern side as well at a slightly smaller radius
$R\sim 3.9\kpc$  (Figure \ref{fig:radialns}b).
The corresponding features can be identified in the LVD (Figure \ref{fig:pv600zoom}c):
around ($l$,$v$)$\sim$(31, 100) for the northern and (343, -75) for the southern side,
though for the northern side the entire wedge at $R=4$-$5\kpc$ has a higher $f_{\rm mol}$.
If these bumps are due to the bar, the enhancement of $f_{\rm mol}$ is only about 20\% there.

Spiral arms and non-circular motions cause secondary variations on the radial decrease
and steepen the radial gradient locally around the tangent radii of the spiral arms (Section \ref{sec:noncirc} and Appendix).
The arrows in Figure \ref{fig:radialns}(a),(b) show the locations of the tangent radii,
and the thick arrows indicate the ones associated with prominent stellar spiral arms.
The radial profiles seem to be steepened around the radii of the Scutum-Crux arm (south)
and possibly the Sagittarius-Carina arm (north), though the global radial trend is still maintained.
No stellar counterparts have been found for the Sagittarius-Carina spiral arm
\citep{Drimmel:2001fk, Benjamin:2005fr, Churchwell:2009vn}.
Non-circular gas motions around this arm may be smaller as they are a response to a stellar spiral potential.

\subsection{Azimuthal Variations}\label{sec:azimuthal}

Azimuthal variations in $f_{\rm mol}$ appear as scatter in Figure \ref{fig:radial} and
\ref{fig:radialns}(a)(b) and
are secondary compared to the dominant radial trend.
The scatter at each radius in these figures represent variations of $f_{\rm mol}$
along the ring at that Galactocentric radius, thereby showing
the range of azimuthal variation.
This measurement is insensitive to the exact locations of spiral arms, and therefore is robust.
In Section \ref{sec:noncirc} and Appendix, we demonstrated that an LVD samples both spiral arm and
interarm regions.
Even though the exact locations of the arms are debatable,
it is certain that some points in Figure \ref{fig:radialns} should represent spiral arms
while the others show interarm regions.
Thus, the scatter indicates the amount of arm/interarm variation.
From Figure \ref{fig:radialns}(c) the RMS scatters are very small:
$\sim3$-12\% at $R<6\kpc$ and 7-16\% outward.
In what follows, we discuss the reasons for the conclusion that the azimuthal variations
are only $\sim20\%$ in the molecule-dominated inner MW  ($R\lesssim6\kpc$; $f_{\rm mol} >50\%$ on average),
while they increase to $\sim40$-$50\%$ at the atom-dominated outskirts ($R\gtrsim6\kpc$; $f_{\rm mol}<50\%$).

Although the scatter plots (Figure \ref{fig:radialns}) show the peak-to-peak variation
for a given $R$, the measurement of average azimuthal variation requires some interpretation.
Figure \ref{fig:radialns} superposes the data from the bar, spiral arms, and interarm regions,
all of which have their own deviations from the circular rotation which smear the plots in the
horizontal direction (Section \ref{sec:noncirc} and Appendix).
Obviously, there must be very localized small regions where stellar feedback dissociates molecules
into atoms. Thus, a peak-to-peak variation at each radius will hide the global trend of
ISM phase change.
In addition, the velocity dispersion of HI gas, $9$-$10\kmps$ \citep{Malhotra:1995lr}, is
larger than that of H$_2$ gas, $3$-$4\kmps$ \citep[cloud-cloud motions; ][]{Clemens:1985lr}.
Therefore, the HI emission is more smoothed out in the velocity direction;
any enhancement of HI emission along spiral arms,
if it exists, is smoothed out into interarm regions.
This would increase the apparent $f_{\rm mol}$ in the spiral arms and
decrease it in the interarm regions. 
Even with these contaminations the radial decrease is very clear, indicating that 
$R$ is the determinant parameter and that the other local variations are only secondary.
From a visual inspection of vertical widths
in Figure \ref{fig:radialns} with these contaminations in mind,
we estimate $\sim20\%$ as the azimuthal, arm/interarm variations
of $f_{\rm mol}$  at $R\lesssim6\kpc$ and $\sim 40$-$50\%$ at the outskirts where $f_{\rm mol}<50\%$.

These arm/interarm variations in the inner MW ($R\lesssim6\kpc$) are much smaller than
those from the complete phase transition suggested by the classic picture of ISM evolution.
The gas remains molecular from interarm regions into spiral arms out again to the next interarm regions.

On the other hand, a rapid phase transition occurs in the outskirts ($R\gtrsim6\kpc$).
For example, the northern side (Figure \ref{fig:radialns}a) shows a few enhancements of $f_{\rm mol}$ around $R\sim 6.0$-$6.5\kpc$,
$6.5$-$7.2\kpc$, and $7.6$-$8.0\kpc$. We can find corresponding enhancements
of $f_{\rm mol}$ in the LVD (Figure \ref{fig:pv600zoom}c)
as two clumps at ($l$,$v$)$\sim$(35, 45) and (40, 33) and a stretch around (22 to 32, 8), respectively.
These appear even more distinct in the CO distribution (Figure \ref{fig:pv600zoom}b)
and therefore are locally dense regions.
Using Figure \ref{fig:pvmodel_width} as a reference for the qualitative locations of the spiral arms,
the first two are probably on the Sagittarius-Carina arm. The third is along the Perseus arm.
In the outskirts of the MW ($R\gtrsim6\kpc$) where the gas is predominantly atomic,
$f_{\rm mol}$ increases by $\sim40$-$50\%$ in the spiral arm when the gas density is
locally high.
This is consistent with the absence of interarm MCs in the outer Galaxy 
where the ambiguity of heliocentric distance is not an issue \citep{Heyer:1998kx, Heyer:1998lr}.

\subsection{Comparisons with Previous Studies}\label{sec:comppast}

\begin{figure}
\epsscale{1.2}
\plotone{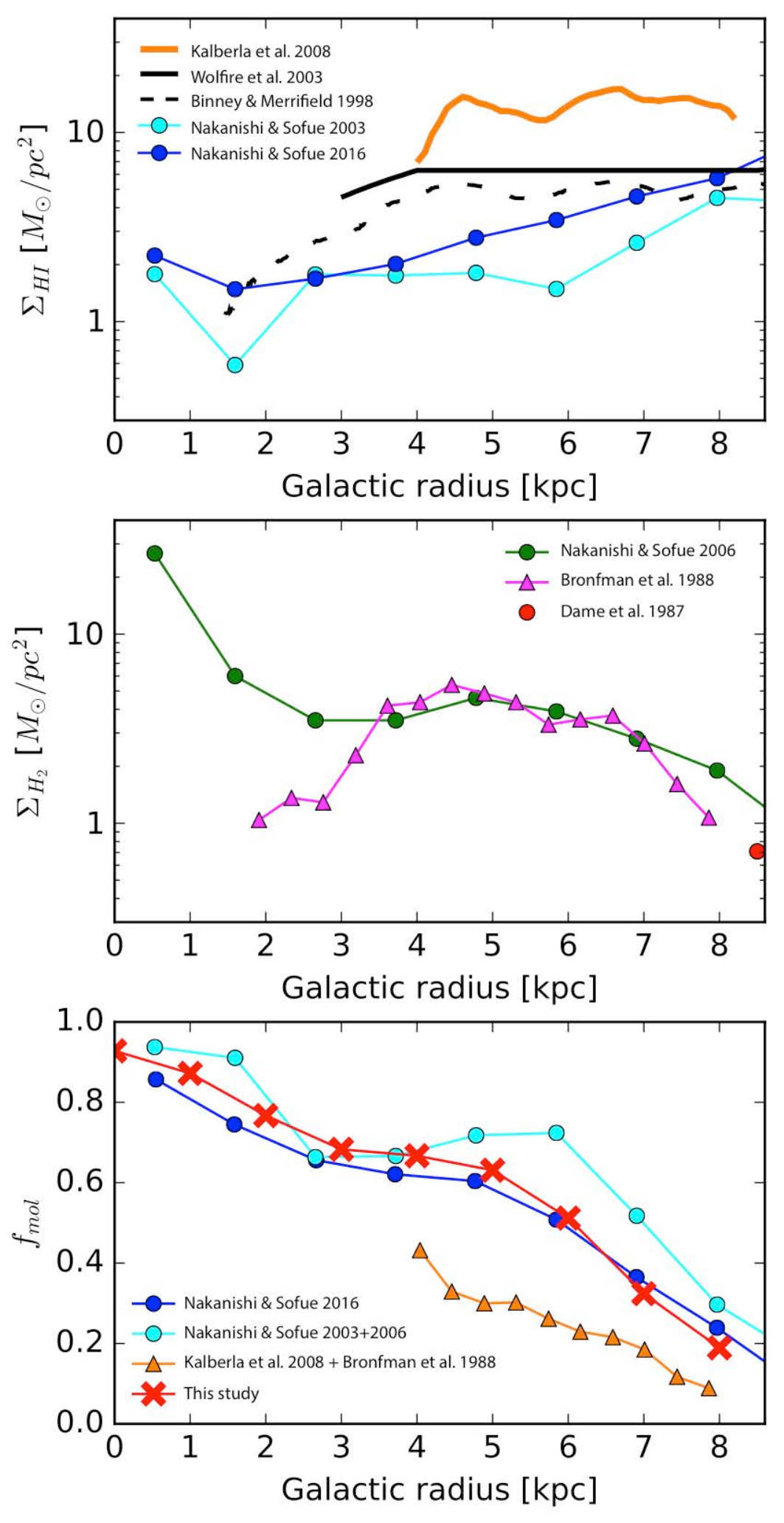}
\caption{
Radial profiles of (a) $\Sigma_{\rm HI}$, (b) $\Sigma_{H_2}$, and (c) $f_{\rm mol}$
derived from the literature for comparison.
The $\Sigma_{\rm HI}$ profiles are from \citet{Kalberla:2008aa}, \citet{Wolfire:2003uq},
\citet{Binney:1998aa}, \citet{Nakanishi:2003xw}, and \citet{Nakanishi:2016aa}.
The data for the first three references are taken from Figure 3 of \citet{Kalberla:2008aa}.
The $\Sigma_{H_2}$ profiles are originally from \citet{Nakanishi:2006uq},
\citet{Bronfman:1988fk}, and \citet{Dame:1987aa}, and the data are taken from \citet{Heyer:2015qy}.
$f_{\rm mol}$ is calculated using the $\Sigma_{\rm HI}$ and $\Sigma_{H_2}$ radial
profiles, except for \citet[][ which calculated $f_{\rm mol}$]{Nakanishi:2016aa} and this study.
All of these studies, except this study, attempted to resolve the ambiguity problem
of kinematic distance, and thus suffer from uncertainties from the problem.
\label{fig:comp}}
\end{figure}

The radial profiles of HI and H$_2$ surface densities, $\Sigma_{\rm HI}$ and
$\Sigma_{H_2}$, have been derived
in previous studies \citep[see reviews, ][]{Kalberla:2009aa, Heyer:2015qy}.
Their measurements in the inner MW are uncertain due to the ambiguity
problem of kinematic distance. Such errors can dilute azimuthal structures, such as
spiral arms, as well as the radial profiles (Section \ref{sec:history} and \ref{sec:circrot}).
Figure \ref{fig:comp} demonstrates this problem and shows
the profiles from the literature.
For example, \cite{Nakanishi:2003xw} and \cite{Kalberla:2008aa}
derived $\Sigma_{\rm HI}$ from the same HI data,
but their results deviate from each other by about an order of magnitude
(e.g., at $R=5$-$6\,\rm kpc$).
\cite{Nakanishi:2016aa} recently revised their analysis,
still the discrepancy remains large.
Their total HI mass within the solar circle is smaller by
a factor of $>2$ than that of \cite{Kalberla:2008aa}.
\cite{Nakanishi:2016aa} suggested that the discrepancy comes
mainly from the adopted vertical profile models.
To separate the gas at near and far sides, these studies fit double Gaussian,
sech$^2$, or more detailed model profiles to emission profiles along $b$.
The high altitude wings of the model profiles may build up and introduce
the large discrepancy in $\Sigma_{\rm HI}$.

The calculations of $\Sigma_{\rm HI}$ and $\Sigma_{H_2}$ radial profiles require heliocentric distances,
and thus a resolution of the near/far ambiguity problem.
Our analysis is less susceptible to this as we do not derive $\Sigma_{\rm HI}$ and $\Sigma_{H_2}$.
Instead, we calculate $f_{\rm mol}$,
a distance independent parameter, in LVD and does not decompose
the gas at near and far sides.
Figure \ref{fig:comp}c compares our $f_{\rm mol}$ profile with those derived
from some combinations of the previous $\Sigma_{\rm HI}$, $\Sigma_{H_2}$ calculations.
The radially-declining trend is common, but the $f_{\rm mol}$ value
varies a lot due to the above difficulties.
In this figure, \citet{Nakanishi:2016aa}
is the closet to our result from the simpler method.
This might indicate that their $\Sigma_{\rm HI}$ and $\Sigma_{H_2}$ calculations,
at least their ratios, are the closest to the reality.

\subsection{Notes on Potential Systematic Errors}\label{sec:syserror}

Potential systematic errors in an LVD analysis have been discussed
\citep[e.g., ][]{Burton:1992aa, Binney:1998aa}. 
Our analysis does not depend on heliocentric distance and
is relatively immune to the near-far distance ambiguity/degeneracy problem.
Nevertheless, there are some potential systematic errors,
most of which we have already discussed.
Here we re-summarize them in terms of three error sources:
(1) the overlap of near and far sides in our analysis,
(2) potential difference in motions of the HI and H$_2$ components, and
(3) possible dark HI and H$_2$ components.

The near-far distance degeneracy could indirectly affect our analysis.
The near and far sides for a given ($l$, $v$) are at the same $R$,
and thus, we analyzed them together.
This treatment may occasionally mix a spiral arm and interarm region
at the near and far sides (Section \ref{sec:method}); in such a case $f_{\rm mol}$ likely
represents the value in the spiral arm as the emission is typically brighter there.
In addition, the fixed beam size and angular scale height over which the emission is
averaged correspond to different physical sizes between near and far sides, and may dilute $f_{\rm mol}$.
This would likely result in an increased scatter, particularly near the Sun ($R\sim 8.5\kpc$),
since a smaller physical size on the near side may pick up local variations,
e.g., inside and outside molecular clouds.
Our averaging scale is large in $b$ (over $60\arcdeg$ or $3\arcdeg$),
but relatively small in $l$ ($0.6\arcdeg$).
In Figure \ref{fig:radial} and \ref{fig:radialns},
some of the scatter around $R\sim 8.5\kpc$  may come from this error.

A difference in the motions of the HI and H$_2$ (CO) components could cause
an additional error. Most likely, the HI gas has a larger velocity width than H$_2$,
which would smear HI spiral arms and leak HI emission from arms into interarm regions.
[The leak from interarms to arms should be smaller as the emission is
concentrated in the arms more than in the interarms.]
This would apparently raise $f_{\rm mol}$ in spiral arms and lower it in
interarm regions, possibly increasing the apparent arm-to-interarm variations
(Section \ref{sec:azimuthal}).
For example, the turbulent velocity dispersion is larger for HI than for H$_2$.
If there are gradients in the rotation velocity v.s. height
(presumably only in the HI layer with a much larger scale hight),
it would also increase the effective velocity width of HI in LVD.

We assumed optically-thin HI 21cm emission and the CO-to-H$_2$
conversion factor for calculations of HI and H$_2$ surface densities.
If optically-thick HI and CO-dark H$_2$ exist, the HI and CO emission
might not accurately trace gas surface densities (Section \ref{sec:caveats}).
These dark HI and H$_2$ should, to an extent, compensate each other in the
$f_{\rm mol}$ calculation.

\section{Discussion}\label{sec:discussion}

\subsection{ISM Evolution in Galaxies}
The azimuthal variation of the ISM phase is an important clue for characterizing ISM evolution
and star formation in galaxies.
In Section \ref{sec:results}, we demonstrated that in the MW the azimuthal variations of the molecular fraction
are much smaller than the radial variations.
In the molecule-dominated inner disk ($f_{\rm mol}>50\%$; $R\lesssim6\kpc$) the gas stays molecular
in both spiral arm and interarm regions. The azimuthal, arm/interarm, variations in $f_{\rm mol}$ are only about 20\%.
In the atom-dominated outskirts ($f_{\rm mol}<50\%$; $R\gtrsim6\kpc$) the variations
can reach as high as 40-50\% in the spiral arms.
The classification of "on-average" molecule-dominated and atom-dominated regions is the key
to understanding the discrepancies in GMC evolution and lifetimes in the literature
\cite[Section \ref{sec:history}; e.g., ][ and see also \citealp{Koda:2013rt}]{Scoville:1979lg, Blitz:1980sh, Cohen:1980ve, Sanders:1985ud}.

In the molecule-dominated disk of M51 the most massive MCs appear exclusively along spiral arms,
while smaller MCs and unresolved molecular emission still dominate over HI in
the interarm regions \citep{Koda:2009wd, Colombo:2014uq}.
The majority of the unresolved emission needs to be in smaller MCs,
since self-shielding is crucial for survival of molecules in the interstellar radiation field
\citep{van-Dishoeck:1988br}.
These considerations suggest the coagulation and fragmentation of molecular gas structures in the spiral arms,
rather than cycling between HI and H$_2$ gas phases.
The massive MCs and their H$_2$ molecules are not fully dissociated into atomic gas,
but are fragmented into smaller MCs on leaving the spiral arms.
The remnants of fragmented massive MCs are detected in the interarm regions
as smaller MCs \citep{Koda:2009wd}.
Dynamical stirring, spiral arm orbit crowding, as well as spiral arm shears, likely
play major roles in the ISM evolution in the molecule-dominated region.
A similar difference in MC mass between spiral arms and interarm regions is found
in the inner MW disk \citep{Koda:2006fk}.
The small azimuthal variations of $f_{\rm mol}$ suggest that
the evolution of the ISM and MCs in the inner MW is similar to the dynamically
driven evolution in M51.

The LMC and M33 are rich in atomic gas, having fewer MCs than the MW and M51
\citep{Fukui:2009lr, Engargiola:2003jo}.
Virtually all of the MCs there are associated with HI spiral arms and filaments.
Molecular emission is absent in the interarm regions.
This distribution indicates a short lifetime for MCs
(i.e., the order of an arm crossing timescale $\sim 30$ Myr).
\citet{Kawamura:2009lr} found a similar lifetime of 20-30 Myr in LMC,
by analyzing the fractions of MCs with and without associated star clusters
and by translating them into the MC lifetime using cluster ages as a normalization.
\citet{Miura:2012yq} also found a similar lifetime of 20-40 Myr in M33.
These short lifetimes appear to be common in the atom-dominated galaxies.
This is consistent with our results for the short lifetime of molecules in the atom-dominated
outskirts of the MW \citep[see also ][]{Heyer:1998kx}.
A similar transition is seen in M51,
whose disk is largely molecule-dominated, but becomes atom-dominated at the very outskirts \citep{Koda:2009wd}.

A transitional case is found in the central $R\sim2$ kpc region of M33 \citep{Tosaki:2011fk}.
The MC distribution is decoupled from the HI structures in the central region \citep{Tosaki:2011fk},
while it coincides with the HI in the atom-dominated outer part \citep{Engargiola:2003jo}.
These decoupled MCs are perhaps entities surviving for long times, greater than
a galactic rotation period during which the HI structures would be smeared out.
$f_{\rm mol}$ increases to $\gtrsim 50\%$ toward the center from $\sim 0$-$20\%$ in the outskirts
\citep[][see their Figure 4]{Tosaki:2011fk}.

\begin{figure}
\epsscale{1.2}
\plotone{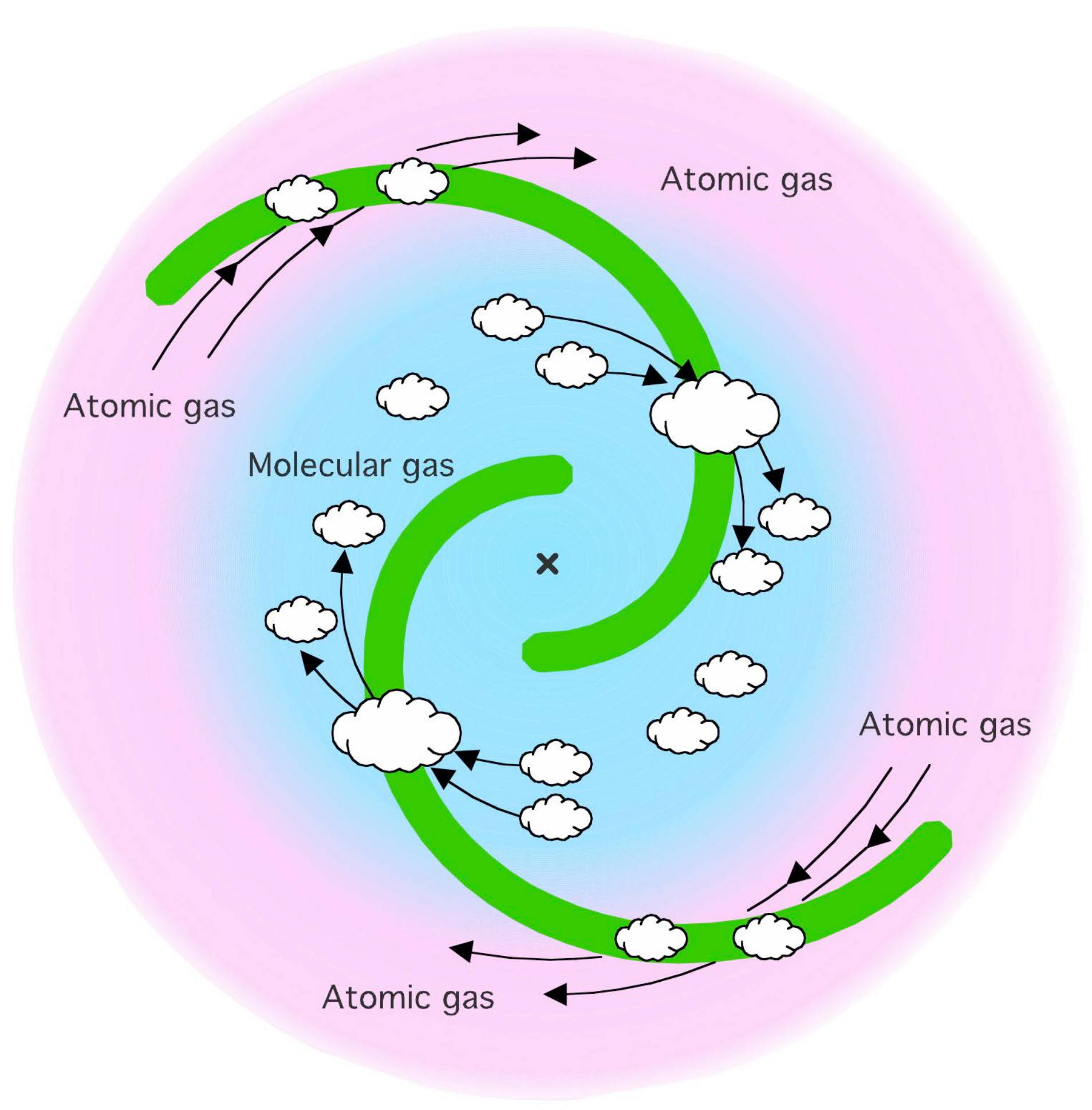}
\caption{Schematic illustration of ISM evolution in the MW.
In the molecule-dominated inner disk (light blue), molecular clouds coagulate into more massive molecular clouds
and then fragment into smaller ones during spiral arm passage. The gas stays in molecular clouds
even in the interarm regions.
On the other hand, in the atom-dominated outer disk (pink), the gas phase transition occurs:
the gas becoming molecular as it enters spiral arms, but being dissociated back into the atomic phase upon exit.
Similar azimuthal evolution is observed in the atom-dominated disks of the LMC and M33, and in the molecule-dominated disk of M51.
[Note that the radial range of this illustration is assumed to be within the co-rotation radius of the MW.]
\label{fig:schema}}
\end{figure}

All of the above point to an integrated view of ISM evolution in galaxies.
In the inner parts of galaxies where the molecular gas is overall dominant,
the gas stays molecular even in the interarm regions.
On the other hand, in the outer atom-dominated parts, the phase transition
occurs in the gas, becoming molecular as it enters spiral arms, but being
photo-dissociated back into the atomic phase upon exit.
Figure \ref{fig:schema} presents a schematic illustration of the ISM evolution in the inner and outer disk.

\subsection{The Azimuthal Constancy and Radial Gradient}

$f_{\rm mol}$ decreases monotonically with Galactic radius,
while its azimuthal variation is small $\sim20\%$.
This suggests that the gas phase balance is approximately
in equilibrium at a given radius, and that
the gas cycling between HI and H$_2$ is in a steady state in the azimuthal direction.
The parameter that controls the phase balance exhibits a strong variation with the radius.
Here we show that a simple model can explain the observed radial trend.
This model is based on one principle and two assumptions.

First, if the ISM continuously cycles the gas between the HI and H$_2$ phases,
the steady state suggests that the HI$\rightarrow$H$_2$ mass conversion rate "$M_{\rm HI}/\tau_{\rm HI}$"
is equal to that for H$_2$$\rightarrow$HI "$M_{\rm H_2}/\tau_{\rm H_2}$" at each radius
\citep[i.e., the continuity principle; ][]{Scoville:1979lg}.
Therefore, we have
\begin{equation}
\frac{M_{\rm H_2}}{M_{\rm HI}}=\frac{\tau_{\rm H_2}}{\tau_{\rm HI}}.\label{eq:cont}
\end{equation}
Based on observations, the mass in the HII phase is taken to be negligible, and the fraction of gas
converted to stars per orbit is small \citep{Bigiel:2008qy}.

Second, we assume that molecules and MCs form from the HI gas exclusively in spiral arms (or in the bar).
Hence the HI$\rightarrow$H$_2$ conversion timescale 
-- or, equivalently, the lifetime of a typical HI atom ($\tau_{\rm HI}$) -- scales with the arm-to-arm travel time.
For a spiral galaxy like the MW,
\begin{equation}
\tau_{\rm HI} = \frac{2 \pi }{\epsilon_{\rm HI} m ( \Omega(R)-\Omega_{\rm p} )},\label{eq:tauHI}
\end{equation}
where $m$ is the number of spiral arms,
$\epsilon_{\rm HI}$ is the HI$\rightarrow$H$_2$ conversion efficiency in a single arm encounter,
$\Omega_{\rm p}$ is the constant pattern speed of the spiral arms,
and $\Omega(R)=V_0/R$ is the angular speed of gas for a flat rotation curve.
$\epsilon_{\rm HI}$ may depend on spiral arm strength and HI density,
e.g., if gravitational collapse followed by spiral arm compression is necessary for the conversion,
but we take it to be constant within $R<R_0$ since the HI density does not vary much
within the solar circle \citep{Burton:1978ve, Scoville:1987vo, Nakanishi:2003xw}.
[Note beyond the co-rotation radius $R_{\rm cr}\equiv V_0/\Omega_{\rm p}$, the sign of eq. (\ref{eq:tauHI})
should be flipped.]

Third, we assume that the H$_2$$\rightarrow$HI conversion timescale -- the lifetime of a typical H$_2$ molecule
($\tau_{\rm H_2}$) -- is constant.
This is justified if the dissociation of molecules and MCs is due to internal physics of the MCs.
For example, if many cycles of star formation are required to completely dissociate
all the molecules in a MC, $\tau_{\rm H_2}$ could be constant in a statistical sense when averaged
over the MC mass spectrum, though it may vary for individual clouds.

Equations (\ref{eq:cont}) and (\ref{eq:tauHI}) give
\begin{eqnarray}
\frac{M_{\rm H_2}}{M_{\rm HI}} &=& \frac{\tau_{\rm H_2} \epsilon_{\rm HI} m}{2\pi} \left( \Omega({R})-\Omega_{\rm p} \right) \\
&=&  \frac{\tau_{\rm H_2} \epsilon_{\rm HI} m \Omega_{\rm p}}{2\pi} \left( \frac{R_{\rm cr}}{R}-1 \right).
\label{eq:modelfmol}
\end{eqnarray}
This model predicts that the molecular fraction decreases with radius.
Qualitatively, $\tau_{\rm HI}$ ($\propto$ the arm-to-arm travel time) increases with increasing Galactic radius,
while  $\tau_{\rm H_2}$ is set to be constant, thus naturally explaining
the transition smoothly from the molecule-dominated inner part to the atom-dominated outer part.

\begin{figure}
\epsscale{1.2}
\plotone{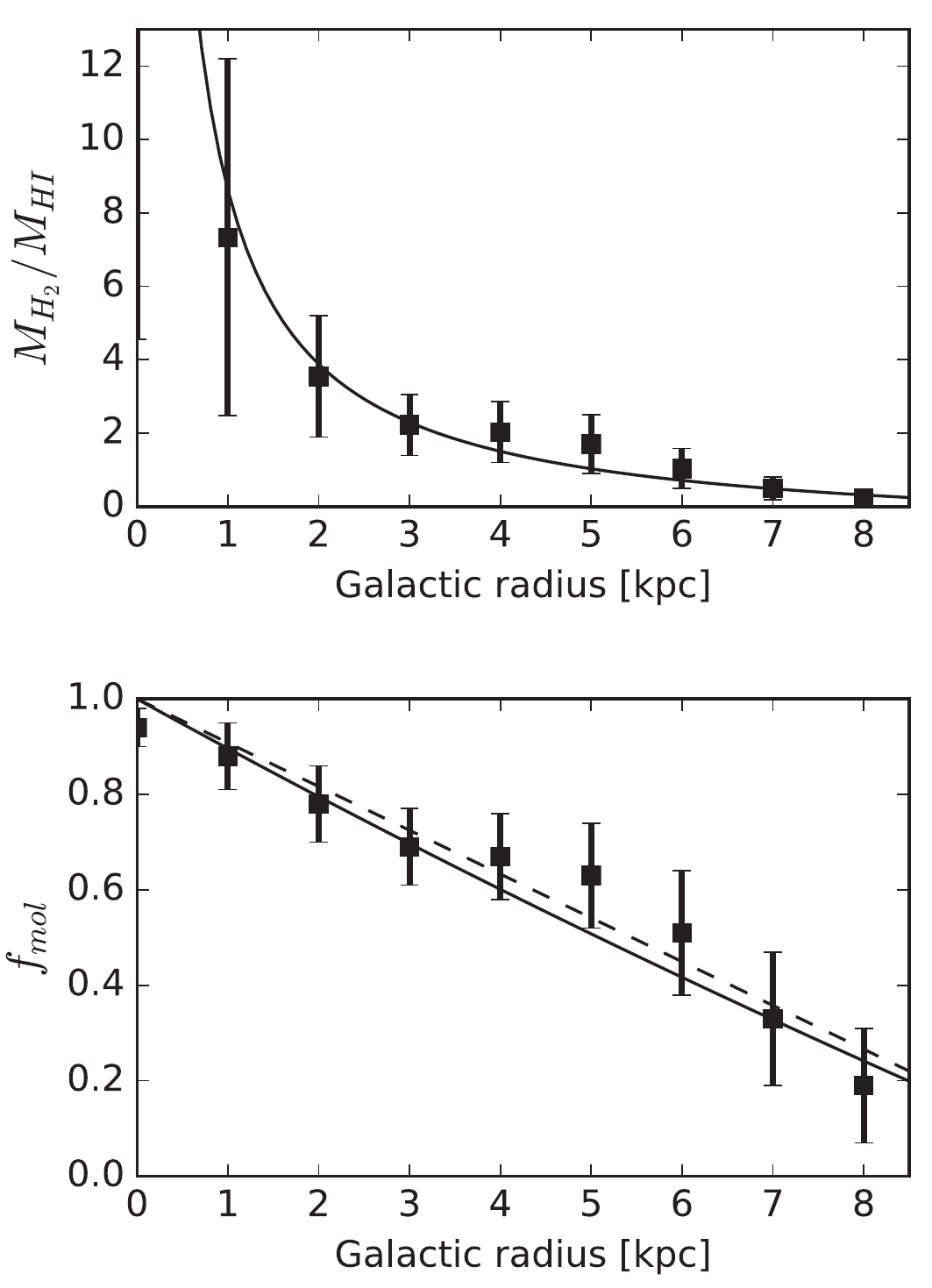}
\caption{
Least-square fits.
The data are the averages of the northern and southern parts and from Table \ref{tab:fmol}.
(a) A fitting function of $M_{\rm H_2}/M_{\rm HI}=\alpha t - \beta$, where $t=1/R$, is used
Note $M_{\rm H_2}/M_{\rm HI}=f_{\rm mol} / (1-f_{\rm mol})$.
The fit provides ($\alpha$, $\beta$) = ($9.5\pm0.5$, $0.87\pm 0.23$) and the solid line.
(b) The same data and fitted line as in (a), but converted to the $f_{\rm mol}$-$R$ plot.
Another fit is also made using a simple linear function,
$f_{\rm mol}=1-\gamma R$. The $y$-intercept is around 1 in the data, so we fix
it to 1. The fit provides $\gamma=0.089\pm0.052$ and is presented by the dashed line.
\label{fig:fmolfit}}
\end{figure}

For quantitative assessment, we make a fit to the data (Table \ref{tab:fmol}) using
a fitting function with the form of eq. (\ref{eq:modelfmol}),
$M_{\rm H_2}/M_{\rm HI} = \alpha t - \beta$,
where $t\equiv 1/R$, $\beta \equiv \tau_{\rm H_2} \epsilon_{\rm HI} m \Omega_{\rm p}/{2 \pi}$,
and $\alpha/\beta \equiv R_{\rm cr}$.
We do not convert this equation to the one for $f_{\rm mol}$ since it is not as simple for fitting purposes.
The data point at $R=0$ ($M_{\rm H_2}/M_{\rm HI}=12.9\pm8.5$) is excluded from the fit because
the ratio diverges there (eq. \ref{eq:modelfmol}).
The fit results in ($\alpha$, $\beta$) = ($0.95\pm0.5$, $0.87\pm 0.23$).
Figure \ref{fig:fmolfit} shows the result of the fit in panel (a) and its conversion to $f_{\rm mol}$ in panel (b).
Clearly, eq. (\ref{eq:modelfmol}) reproduces the observed radial trend well.

This result translates to $R_{\rm cr}=10.9\pm2.9$ kpc,
$\Omega_{\rm p}=V_0/R_{\rm cr} = 20.1\pm5.3$ km/s/kpc,
$\tau_{\rm H_2}=2 \pi \beta / (\epsilon_{\rm HI} m \Omega_{\rm p})=(132\pm49)\epsilon_{\rm HI}^{-1}(2/m)$ Myr.
These are consistent with those ($\Omega_{\rm p}$, $R_{\rm cr}$)$\sim$(18.4 km/s/kpc, 11.9 kpc)
derived, e.g., by \citet{Bissantz:2003lr} after correction for the adopted $R_0$,
though all measurements in the literature have considerable uncertainties.
The MW has two stellar spiral arms, and therefore, $m=2$.
We estimate $\epsilon_{\rm HI}\sim 0.4$ by translating the $\sim 20\%$ azimuthal variation of $f_{\rm mol}$
where the HI fraction is about 50\% ($f_{\rm mol} = 0.5$; Section \ref{sec:azimuthal}).
Hence $\tau_{\rm H_2} \sim 330$ Myr on average within the solar circle.
Table \ref{tab:modelfit} summarizes the derived parameters.

$\tau_{\rm HI}$ is roughly comparable to the gas rotation timescale with respect to the spiral 
pattern when $\epsilon_{\rm HI} m \sim 1$ (eq. \ref{eq:tauHI}) as in our case,
and $\tau_{\rm HI}\sim$40, 140, 320, 670, and 1700 Myr at $R=$1, 3,  5,  7, and 9 kpc, respectively.
The lifetime of H$_2$ is longer than the rotation timescale in the inner MW,
and the gas stays mostly molecular during the arm-to-arm travel time.
[This is true even if the number of spiral arms is assumed to be $m=4$: $\tau_{\rm H_2}$ ($\propto 1/m$)
would be twice shorter, but the arm-to-arm travel time is also twice shorter.]
The opposite is the case in the outskirts, where molecules survive for only a small fraction of
the rotation timescale and exist only around spiral arms.
Indeed, there are molecular clouds around spiral arms in the outer MW
with $R>R_0$ \citep{Heyer:1998lr, Heyer:1998kx}, but averaged along annulus, $f_{\rm mol}\ll1$.

Some of the parameters may vary in other environments.
For example, $\tau_{\rm HI}$ could be longer
in the outer MW and in the atom-dominated galaxies (e.g. the LMC and M33),
because of an absence of (or weaker) stellar spiral structures (i.e., smaller $m$ and/or lower $\epsilon_{\rm HI}$),
and because of the intrinsically low gas density (i.e., lower $\epsilon_{\rm HI}$
-- at a low density, spiral arm compression, when it exists, may not convert HI to H$_2$ efficiently).
$\tau_{\rm H_2}$ could be smaller if the average MC mass is lower.
All of these keep $M_{\rm H_2}$ and $f_{\rm mol}$ lower
and qualitatively explain the HI-dominated regions.

We assumed that spiral arms or a bar are the trigger of the HI$\rightarrow$H$_2$ conversion,
however, this model works even if the conversion is due to another physical mechanism
as long as the timescale is comparable/proportional to the rotation timescale at that radius.
For example, if MCs form by the agglomeration of atomic clouds and
if their collision timescale is set by their velocity difference due to differential
galactic rotation, $\tau_{\rm HI}$ would have a similar dependence on $\Omega(R )$
\citep[e.g., ][]{Scoville:1979lg, Wyse:1986fk, Wyse:1989qy, Tan:2000lr}.
This model assumed that the arms/bar enhance the HI$\rightarrow$H$_2$ conversion significantly.
We should note that the conversion could occur at a much
lower rate, e.g., in a dwarf galaxy without prominent spiral arms/bar,
if there are local density fluctuations. 

\begin{deluxetable}{lcccc}
\tablecolumns{5}
\tablewidth{0pc}
\tablecaption{Parameters from fit}
\tablehead{
\colhead{} & \colhead{} & \multicolumn{3}{c}{$X_{\rm CO}$ [${\rm cm^{-2}\,(K\,\kmps)^{-1}}$]} \\
\cline{3-5}
\colhead{Parameter} & \colhead{Unit}   & \colhead{$2\times 10^{20}$}  & \colhead{$3\times 10^{20}$}}
\startdata
$\alpha$                                 &  & $9.5\pm0.5$& $14.3\pm0.8$ \\ 
$\beta$                                    & & $0.87\pm0.23$& $1.31\pm0.34$ \\ 
$R_{\rm cr}$ ($=\alpha/\beta$)  & kpc & $10.9\pm2.9$& $10.9\pm2.9$\\ 
$\Omega_{\rm p}$ ($=V_0/R_{\rm cr}$)  & km/s/kpc & $20.1\pm5.3\tablenotemark{a} $& $20.2\pm5.3$\tablenotemark{a} \\ 
$\tau_{\rm H_2}$ ($=2\pi \beta / m \epsilon_{\rm HI} \Omega_{\rm p}$) É & Myr & $331\pm 123$\tablenotemark{b} & $499\pm184$\tablenotemark{b}

\enddata
\tablenotetext{a}{Assumed $V_0=220$ km/s.}
\tablenotetext{b}{Assumed $\epsilon_{\rm HI}=0.4$ and $m=2$.}
\label{tab:modelfit}
\end{deluxetable}

\subsection{Comments on the Midplane Pressure}
The radial decrease of $f_{\rm mol}$ alone has been known for some time  \citep{Sanders:1984fx, Young:1991kq, Sofue:1995fk, Honma:1995qy, Wong:2002lr}
though the azimuthal variation was rarely analyzed.
This radial trend is often discussed in relation to the hydrostatic pressure at the midplane of
galactic disk under the gas and stellar gravitational potentials, $P_{\rm mid}$
(\citealt{Wong:2002lr, Blitz:2004zn, Blitz:2006or, Field:2011lr, Hughes:2013lr};
see also \citealt{Elmegreen:1993fk, Elmegreen:1989lr}).
There is an empirical linear correlation between $P_{\rm mid}$
and the mean ratio of molecular to atomic hydrogen $R_{\rm mol}$.
However, it remains unclear how $P_{\rm mid}$ is physically coupled to the MCs, the major reservoir of molecular gas.

It is often mistakenly assumed that the ambient midplane pressure confines the gas in MCs.
This is not the case, and the nature of the pressure needs to be considered carefully.
The thermal and magnetic pressures are not strong enough to confine the gas within a MC,
while the pressure from large-scale turbulence is not a confining pressure.
Adopting a supersonic dispersion of $3$-$5\kmps$,
the internal turbulent pressure of MCs is $P_{\rm MC}/k \sim 10^{5-6}\,\rm K\cdot cm^{-3}$.
This exceeds the thermal pressure of the ambient gas, $P_{\rm th}/k\sim 10^{3-4}\,\rm K\cdot cm^{-3}$,
by 1-2 orders of magnitude.
The magnetic pressure is also too low, $P_{\rm B}/k = B^2/(8\pi k)\sim10^{4}\,\rm K\cdot cm^{-3}$
using the observed magnetic strength of $B<10\mu G$ in the ambient medium \citep{Crutcher:2012lr}.
The external turbulent pressure $P_{\rm turb}$ which supports the vertical structure
does not confine the gas within MCs,
since it is mostly unisotropic/directional -- a MC may feel ram pressure from the direction
of its motion with respect to the ambient gas, but this head wind is only from one side of the MC,
and there is no turbulent pressure on its trailing side.
The midplane pressure cannot directly confine gas in MCs.

A careful assessment of causalities and physical mechanisms for the quasi-equilibrium
of the gas phases is needed. \citet{Ostriker:2010lr} made such an attempt and distinguished
$P_{\rm th}$ from $P_{\rm mid}$ ($=P_{\rm turb}+P_{\rm th}$), but had to assume that the $P_{\rm mid}$ from
the vertical dynamical equilibrium and the $P_{\rm th}$ for the thermal equilibrium
are coupled (they assumed $P_{\rm mid}/P_{\rm th}=5$).
The establishment of such energy partition -- from the galaxy center to outskirts and between
spiral arms and interarm regions --
is the key question to understanding the phase balance in the ISM.
In addition to the complex energy balance in the ISM,
many parameters have radial dependences and are inter-dependent.
For example, $P_{\rm mid}$ is often calculated from stellar and gas surface densities alone,
and one should ask, e.g., which parameter really causes the phase balance (pressure or density?).
Future studies should carefully sort out these degeneracies.

\section{Conclusions}

We analyzed the variations of molecular fraction $f_{\rm mol}$ in the MW in the radial and azimuthal
directions by using the archival CO($J$=1-0) and HI 21cm emission data.
$f_{\rm mol}$ decreases monotonically from
the globally molecule-dominated central region ($f_{\rm mol}\sim 100\%$) to
the mostly atom-dominated outer region of the Milky Way
($f_{\rm mol}\sim 10$-$20\%$ at the Solar radius when integrated over the whole gas disk
thickness $|b|<30\arcdeg$ and $\sim 50\%$ at the disk midplane).
The azimuthal variation, and hence arm/interam variation, of the gas phase
is small, $\sim 20\%$, within the molecule-dominated inner disk
($R\lesssim6\kpc$; $f_{\rm mol}\gtrsim 50\%$).
The gas stays largely molecular even after spiral arm passage and in interarm reigons.
This is at variance with the classic scenario of ISM evolution for rapid and complete
phase transitions during spiral arm passage.
On the contrary, the rapid gas phase change occurs only in the atom-dominated outskirts
($R\gtrsim6\kpc$; $f_{\rm mol}\lesssim 50\%$).
The average $f_{\rm mol}$ around the solar neighborhood is about 20\% including
the HI gas at high Galactic disk altitudes, while it is still $\sim$50\% 
at the disk midplane at the solar radius.
The gas stays largely molecular across spiral arms and interam regions in its inner disk,
while in the outskirts the molecular gas is localized
in the spiral arms and becomes atomic in the interarm regions.
This classification of on-average atom-dominated and molecular-dominated regions
appear to be applicable to other nearby galaxies, such as LMC, M33, and M51.

We also demonstrated that a simple model of the phase balance and mass continuity
in the HI and H$_2$ cycling can explain the observed radial trend,
if the HI$\rightarrow$H$_2$ conversion occurs on a galactic rotation timescale (e.g, 
due to spiral arm compressions) and the H$_2$$\rightarrow$HI conversion has
a constant timescale (e.g., due to internal physics of molecular clouds, such
as multiple cycle of star formation).

\acknowledgments
We thank Roberta Paladini, Tom Dame, Lena Murchikova, and Jim Barrett for useful discussions.
We also thank the anonymous referee for careful reading.
JK thanks California Institute of Technology for hospitality during extended visits.
This work is supported by the NSF through grant AST-1211680. JK also acknowledges the supports from NASA
through grants NNX09AF40G, NNX14AF74G, a Herschel Space Observatory grant, and an Hubble Space Telescope grant.

\clearpage
\appendix

\section{Milky Way Spiral Arms in a Longitude-Velocity Diagram}\label{sec:modelvld}
A longitude-velocity ($l$-$v$) diagram (LVD) is a tool to investigate spiral arms and interarm regions
in the Milky Way. Here we demonstrate how spiral arms appear in an LVD using an observationally-motivated,
but simplistic, logarithmic spiral model.

\subsection{Spiral Arms in the Milky Way}
The MW is likely to have two major, and potentially two minor, spiral arms
\citep{Drimmel:2000lr,Benjamin:2008ul,Churchwell:2009vn, Robitaille:2012lr}.
\cite{Steiman-Cameron:2010lr} obtained the geometry of the four spiral arms in the gas component
using the longitude profile of [CII] 158$\micron$ line emission. They assumed that each emission peak
indicates the longitude of a spiral arm tangent point, and that two tangential longitudes of a spiral arm determine
the geometry of the arm.
Figure \ref{fig:model1}a  shows the face-on projection of the four gas arms.
Only two of the four arms were identified in stellar distributions
\citep{Drimmel:2000lr,Benjamin:2008ul,Churchwell:2009vn}.
\citet{Robitaille:2012lr} concluded that a model with two major and two minor spiral arms can
reproduce the range of emission from stellar to dust infrared, and to polycyclic aromatic hydrocarbon (PAH) emission.
The MW has a relatively large bar at the center \citep{Benjamin:2005fr}, and
external galaxies with such large bars most often have only two significant stellar spiral arms.
At the same time, optical images of a number of barred galaxies, e.g., from the Hubble Space Telescope archive,
show filamentary dust-extincted lanes between stellar spiral arms. 
These ISM concentrations would appear as apparent spiral arms if observed in tracers of the ISM and associated
star formation in an LVD.
As per historical convention, we call the four gas arms the Sagittarius-Carina, Scutum-Crux,
Norma-Cygnus, and Perseus arms. The Perseus and Scutum-Crux arms are 
the stellar spiral arms (thick lines in Figure  \ref{fig:model1}a).

\subsection{A Simple Model of Spiral Arms in A Longitude-Velocity Diagram}
\begin{figure*}
\epsscale{1.0}
\plotone{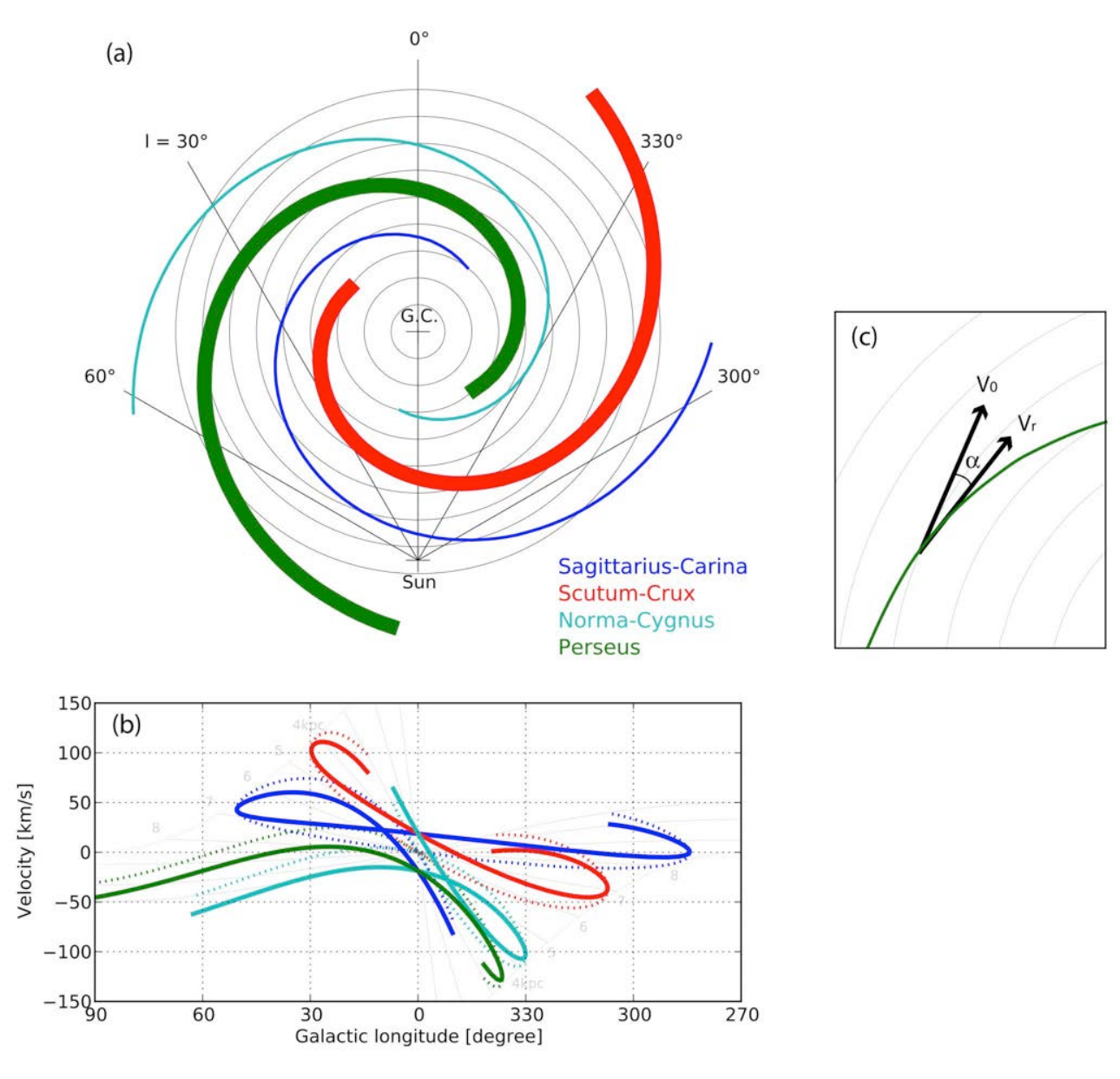}
\caption{Logarithmic spiral arms of the Milky Way.
(a) Logarithmic spiral arms from \cite{Churchwell:2009vn, Steiman-Cameron:2010lr}.
Four arms, Sagittarius-Carina (blue), Scutum-Crux (red), Norma-Cygnus (cyan), and Perseus arms (green), are often
identified by studies with ISM and star formation tracers while only the Perseus and Scutum-Crux arms (thick lines)
are identified as stellar spiral arms.
Concentric circles around the Galactic Center (G.C.) are drawn at a 1-kpc interval.
(b) Longitude-velocity diagram of the four spiral arms. Dotted lines are for the model of pure circular rotation with a constant rotation velocity
of $220\kmps$ and the solar Galactcentric radius of $R=8.5\kpc$. Solid lines are for the model with non-circular motions, $\alpha=5\arcdeg$
and $V_r=215\kmps$.
(c) Explanation of the tilt angle $\alpha$ measured from the pure circular rotation vector $V_0$ to the non-circular motion vector $V_r$.
\label{fig:model1}}
\end{figure*}

These spiral arms are translated into the $l$-$v$ space using a model of MW rotation.
\cite{Pineda:2013lr} assumed a pure circular flat rotation curve with a velocity of
$V_0=220\kmps$ and the Sun at a radius of $R_0=8.5\kpc$.
At a general location in the MW disk, its Galactic radius $R$ and longitude $l$
determine an observed line-of-sight velocity $v_l$ as
\begin{equation}
v_l = V_0 \left( \frac{R_0}{R} -1 \right) \sin l,
\end{equation}
which is the same as eq. (\ref{eq:rotv}).
Figure \ref{fig:model1}b (dotted lines) shows the spiral arms with the circular rotaiton on an LVD.

Non-circular motions, i.e., deviations from the pure circular rotation, change the arm locations on the LVD.
The gas and stars take elongated (oval) orbits due to the kinematic density wave \citep{Onodera:2004pb}.
They slow down and stay long around the apocenter, since it's the outermost radius of the orbit in the Galactic
gravitational potential. This slow-down causes an enhanced density in the spiral density wave.
\citet{Koda:2006th} demonstrated this density enhancement in the case of a bar potential,
and the same mechanism should work in a spiral potential \citep[see ][]{Onodera:2004pb}.
Thereby, the rotation velocity $V_r$ with non-circular motions on a spiral arm (i.e., apocenter) should
be smaller than that with pure circular rotation $V_0$.
The direction of motion should also be tilted slightly inward toward the Galactic center,
by a small angle $\alpha$, with respect to the tangential direction of the circular orbit.
Figure \ref{fig:model1}c shows the definitions of $V_r$ and $\alpha$.
In this case, the spiral arm is expressed as,
\begin{equation}
v_{\rm l} = V_r \left[ \left( \frac{R_0}{R} \right) \sin l \cos \alpha \pm \sqrt{1-\left( \frac{R_0}{R} \sin l\right)^2 \sin \alpha} \right] \\
- V_0 \sin l.
\end{equation}
The "$+$" is for the far side and "$-$" for the near side 
since a line-of-sight typically passes a single spiral arm twice (see Section \ref{sec:circrot} for near/far distances).
The Sun is assumed to be on a circular orbit.
We should note that this expression is only for the points on spiral arms, not for other parts of the orbit.

Figure \ref{fig:model1}b (solid lines) shows the spiral arms with the non-circular motions.
We arbitrarily assumed a constant $\alpha=5\arcdeg$ and $V_r=215\kmps$ for all Galactcentric radii $R$.
The arms form coherent loops as in Figure \ref{fig:model1}b, and the interarm regions
appear between the spiral arm loops.
The purpose of this model is only the qualitative demonstration of the effects of non-circular motions
around spiral arms on an LVD.
$\alpha$ and $V_r$ should, of course, vary with radius and could be different between the spiral arms.
Nevertheless, this model appears closer to the spiral arms traced by the distribution of HII regions
in an LVD \citep{Sanders:1985ud} than does a model with only pure circular rotation.

\subsection{Effects of Spiral Arms and Non-Circular Motions in a Longitude-Velocity Diagram}
This toy model provides an insight into how spiral arms and interarm regions should appear
in an LVD and how they affect our analysis. Two points are important.
First, the locations of spiral arms systematically shift in the $v_l$ direction in an LVD due to non-circular motions,
which cause errors in $R$.
Second, even with increased velocity widths due to enhanced velocity dispersions and spiral arm
streaming motions, interarm regions are still sampled in an LVD.

\begin{figure*}
\epsscale{1.1}
\plotone{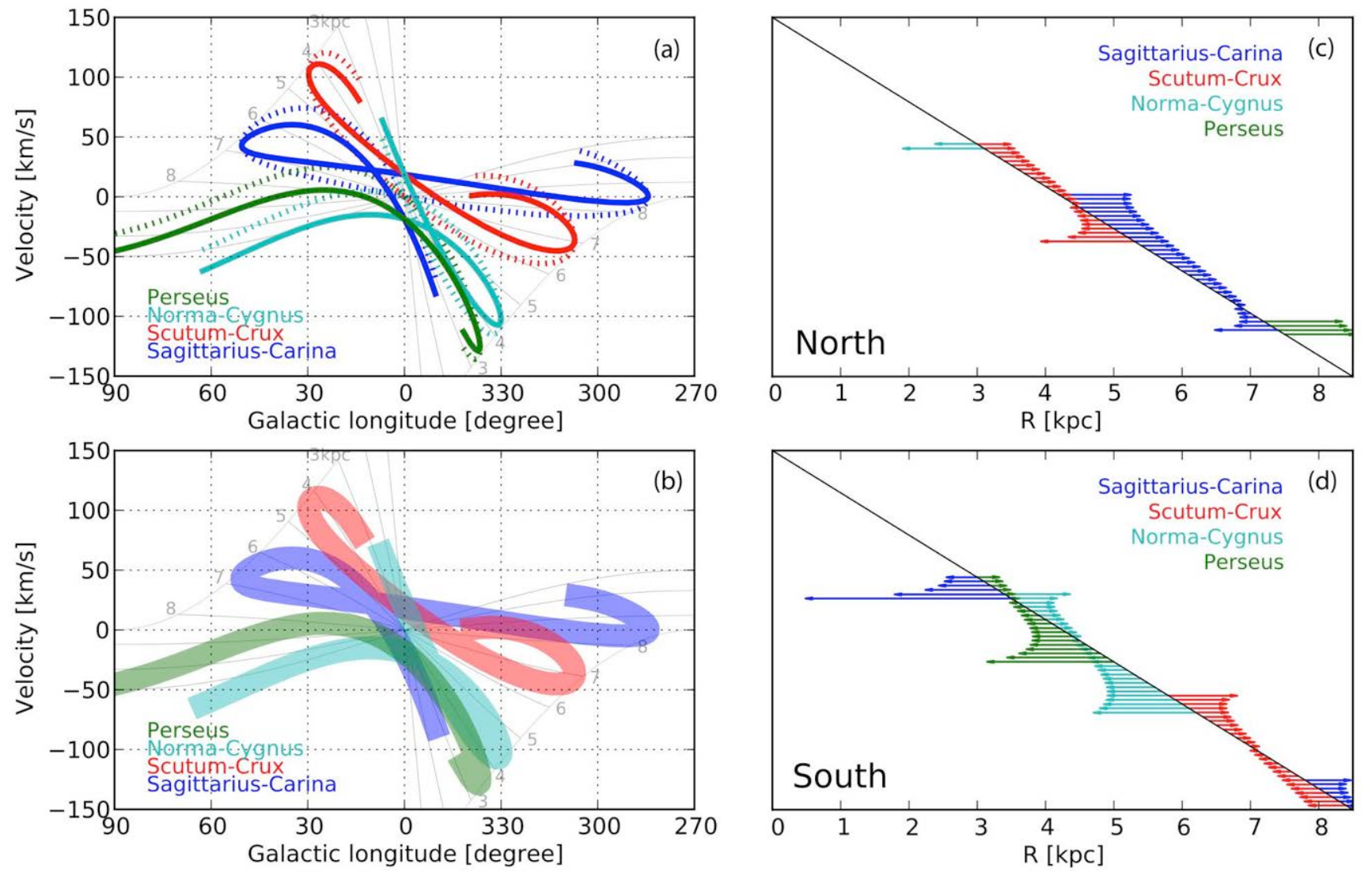}
\caption{Model $l$-$v$ diagrams.
(a) The same as Figure \ref{fig:model1}b.
Four spiral arms are plotted in the cases of pure circular rotation (dotted lines) and
with non-circular motions (solid lines).
(b) Four spiral arms with non-circular motions. The widths of the lines correspond to $20\kmps$ (arbitrarily chosen)
to demonstrate the effect of velocity dispersions and streaming motions in spiral arms. The actual widths are likely
smaller by a factor of about 2, especially for the Sagittarius-Carina and Norma-Cygnus arms that have no enhancement
in stellar density (potential).
(c)(d) Systematic errors in $R$ at positions along spiral arms due to non-circular motions:
(c) for the northern, $0\arcdeg< l < 90\arcdeg$ and $v>0\kmps$, and
(d) southern disk, $270\arcdeg< l < 360\arcdeg$ and $v<0\kmps$.
The arrows indicate the amounts of shifts from the true radii, i.e., the origins of the arrows.
For example, if an underlying radial profile follows the black solid line, the profile would be steepened locally
due to the non-circular motions (by connecting the arrowheads).
For clarity, we applied eq.(\ref{eq:removal}) and removed a couple of outermost arrows from the southern parts
of Norma-Cygnus and Perseus arms which are the points around ($l$,$v$)$\sim$(0,0) in panel (a)(b).
\label{fig:pvmodel}}
\end{figure*}

Figure \ref{fig:pvmodel}a demonstrates the systematic shifts of spiral arm locations from the circular rotation
model (dotted lines) to the non-circular motion model (solid).
Spiral arms show loops in the LVD and  tend to appear squashed in the velocity direction
due to the non-circular motions.
This squashing is primarily due to the systematic change of velocity vector directions with respect
to the directions of our lines-of-sight.
For example, if the northern side of the Scutum-Crux arm is considered (Figure \ref{fig:model1}a; $l\sim0$-$30\arcdeg$),
the velocity vectors at its far side rotate away from our lines-of-sight due to the non-circular motions,
while those at the near side rotate toward them.
This shows that the squashed spiral loops are a general consequence of spiral arm non-circular motions
in the MW.
 
The shifts in arm locations in an LVD result in systematic shifts in $R$ when equation (\ref{eq:rotv}) is used.
Figure \ref{fig:pvmodel}a also shows constant $R$ lines.
Figure \ref{fig:pvmodel}c,d qualitatively demonstrate how these shifts affect radial profiles,
separately for the northern and southern sides of the MW disk.
The arrows indicate the directions and (very roughly) amounts of systematic shifts in $R$.
If an underlying radial profile of, e.g., $f_{\rm mol}$ follows the black solid line,
the profile would shift in the directions of the arrows.
As a result, it would appear steepened (locally in some $R$ range) due to
non-circular motions.
The degree of steepening, of course, depends on that of non-circular motions.

The effect of increased velocity widths is demonstrated in Figure \ref{fig:pvmodel}b.
The width is manually set for the non-circular rotation model;
it corresponds to $\sim20\kmps$ when an arm runs horizontally in the LVD, and is even wider,
up to $\sim 50\kmps$, around the tangent points of the spiral arms.
Even such widened spiral arms do not completely fill the LVD.
Therefore, the LVD samples both spiral arm and interarm regions, even though
their true locations remain uncertain.
The actual widths of spiral arms are likely about twice as narrow in the velocity domain,
and this figure likely shows their largest possible impact on the LVD.
In fact, most observations of spiral arm streaming motions and velocity dispersions in the MW indicate
a smaller full width \citep[$2$-$12\kmps$; e.g., ][]{Clemens:1985lr, Alvarez:1990qf, Oka:2007vn}.
In addition, the Sagittarius-Carina and Norma-Cygnus arms do not show corresponding stellar spiral arm
potentials \citep{Drimmel:2001fk, Benjamin:2008ul, Robitaille:2012lr}, and the steaming motions
are perhaps smaller around these arms.
Spiral arm velocity widths appear to be $<20\kmps$ in numerical simulations of MW gas dynamics 
\citep{Wada:1994ys, Fux:1999yq, Bissantz:2003lr, Pettitt:2015gf}.

\end{document}